\documentclass[aps,prl,twocolumn,superscriptaddress,showpacs,final,floatfix,longbibliography]{revtex4-2}

\usepackage[utf8]{inputenc} 
\usepackage[toc,page]{appendix}
\usepackage{amsfonts}
\usepackage{color}
\usepackage{amsmath}
\usepackage{graphicx}
\usepackage{amssymb}
\usepackage{hyperref}
\usepackage{color}
\usepackage{mathrsfs}
\usepackage{isomath}
\usepackage{amsthm}
\usepackage{epstopdf}
\usepackage{txfonts}
\usepackage{dsfont}
\usepackage{ulem}
\usepackage{physics} \usepackage{tikz} 
\allowdisplaybreaks[4]
\usepackage{bbold}
\usepackage{mathtools}
\usepackage{cleveref}
\usepackage{subfigure}

\newcommand{\ignore}[1]{}

\newcommand{\nnorm}{\nu}

\newcommand{\de}{{\mathrm d}}

\renewcommand{\vec}[1]{\boldsymbol{#1}}

 \renewcommand{\tr}{\mathrm{tr}}

\newcommand{\be}{\begin{equation}}
\newcommand{\ee}{\end{equation}}
\newcommand{\eea}{\end{eqnarray}}
\newcommand{\bea}{\begin{eqnarray}}
\renewcommand{\var}[1]{\ensuremath{(\Delta #1)^2}}
\newcommand{\av}[1]{\ensuremath{\langle{#1} \rangle}}

\renewcommand{\vec}[1]{\boldsymbol{#1}}

\newcommand{\diag}{{\rm diag}}

\newcommand{\id}{\mathbb{1}}

\newcommand{\C}{X_\varrho}

\newcommand{\SN}{\mathcal{SN}}

\newcommand{\intdalpha}[1]{\int_{\mathbb C} \ #1 \ \mathrm{d}^{2} \alpha}

\newtheorem{lemma}{Lemma}

\newtheorem{proposition}{Proposition}

\usepackage{cleveref}
\crefname{equation}{Eq.}{Eqs.}
\crefname{figure}{Fig.}{Figs.}
\crefname{table}{Table}{Tables}
\crefname{observation}{Obs.}{Obs.}
\crefname{corollary}{Corollary}{Corollaries}
\crefname{lemma}{Lemma}{Lemmata}
\crefname{proof}{Proof}{Proofs}
\creflabelformat{proof}{#2proof#3}
\crefname{remark}{Remark}{Remarks}
\crefname{prop}{Proposition}{Propositions}

\begin{document}

\title{Entanglement Dimensionality of Continuous-Variable States From Phase-Space Quasi-Probabilities}

\author{Shuheng Liu}
\affiliation{State Key Laboratory for Mesoscopic Physics, School of Physics, Frontiers Science Center for Nano-optoelectronics, Peking University, Beijing 100871, China}
\affiliation{Vienna Center for Quantum Science and Technology, Atominstitut, TU Wien, 1020 Vienna, Austria}

\author{Jiajie Guo}
\affiliation{State Key Laboratory for Mesoscopic Physics, School of Physics, Frontiers Science Center for Nano-optoelectronics, Peking University, Beijing 100871, China}

\author{Matteo Fadel}
\email{matteofadel@pku.edu.cn}
\affiliation{Institute of Quantum Science and Technology $\&$ Frontiers Science Center for Nano-Optoelectronics, School of Physics, Peking University, Beijing 100871, China}
\affiliation{Department of Physics, ETH Z\"{u}rich, Z\"{u}rich 8093, Switzerland}

\author{Qiongyi He}\email{qiongyihe@pku.edu.cn}
\affiliation{State Key Laboratory for Mesoscopic Physics, School of Physics, Frontiers Science Center for Nano-optoelectronics, Peking University, Beijing 100871, China}
\affiliation{Collaborative Innovation Center of Extreme Optics, Shanxi University, Taiyuan, Shanxi 030006, China}
\affiliation{Hefei National Laboratory, Hefei 230088, China}

\author{Marcus Huber} \, \email{marcus.huber@tuwien.ac.at}
\affiliation{Vienna Center for Quantum Science and Technology, Atominstitut, TU Wien,  1020 Vienna, Austria}
\affiliation{Institute for Quantum Optics and Quantum Information (IQOQI), Austrian Academy of Sciences, 1090 Vienna, Austria}

\author{Giuseppe Vitagliano}\email{giuseppe.vitagliano@tuwien.ac.at}
\affiliation{Vienna Center for Quantum Science and Technology, Atominstitut, TU Wien, 1020 Vienna, Austria}
\affiliation{Instituto de F\'isica Corpuscular, Universitat de Val\'encia - Consejo Superior de
Investigaciones Cient\'ificas, Parc Cient\'ific, E-46980 Paterna, Valencia, Spain}

\begin{abstract}
\vspace{0pt}
The dimensionality of entanglement, quantified by the Schmidt number, is a core resource for quantum information processing. While its certification is well-established in finite-dimensional systems, many platforms harnessing high Schmidt numbers are natively continuous-variable (CV) systems, which are usually truncated to a finite subspace before canonical discrete-variable techniques are applied. For those instances, a core question is whether directly utilizing a genuinely CV toolbox leads to better and more robust characterisations. We answer this question by introducing Schmidt-number witnesses built directly from correlations and covariances of phase-space observables, accessible through characteristic- or Wigner-function measurements. They hold for arbitrary mixed states, including non-Gaussian ones, and only require correlations along a slice of phase space. We show that this direct estimation is more robust and versatile than first discretizing the system and using canonical discrete variable techniques. Moreover, we apply our method to certify arbitrarily-large Schmidt numbers of a bound-entangled Gaussian state. These results motivate genuine CV methods for characterizing high-dimensional entanglement across CV quantum technologies.
\end{abstract}

\maketitle

{\bf Introduction.---}
Significant effort has been devoted to characterizing entanglement in continuous-variable (CV) quantum systems \cite{BraunsteinQuantum2005,WeedbrookGaussian2012} in experimentally accessible ways. Most prominently, conditions on the covariance matrix of the quadratures have been investigated \cite{SimonPeresHorodecki2000,DuanInseparability2000}, as well as approaches based on the moment of the mode operators \cite{ShchukinInseparability2005,HilleryEntanglement2006,MiranowiczInseparability2009,GiedkeEntanglement2001,vanLoockDetecting2003,hyllus2006optimal,ZhangGenuine2023}.
Other approaches are instead based on phase-space quasi-probabilities, which can be seen as arising from correlations of CV {\it local orthogonal observables} (CVLOOs), such as displacements or displaced-parity operators \cite{YuSixiaEntanglement2005,ZhangDetectingPRL2013}.
From these methods, it is also possible to quantify entanglement, e.g., by bounding the entanglement of formation or the negativity \cite{vidal2000entanglement,QuantumHorodecki2009,hyllus2006optimal,ZhangDetectingPRL2013,SchneelochQuantifying2019}. 
So far, however, they are not tailored to bound the Schmidt number, also called {\it entanglement dimensionality}.

This can be singled out amongst entanglement monotones in various contexts~\cite{FriisNatPhys19,SperlingTheSchmidtNumber2011,BaeMoreEntanglementPRL2019,BuscemiEntanglementCostPRL2011,SanperaSchmidtNumberPRA2001,KraftCharacterizingGenuinePRL2018}, and it plays a central role in an ever increasing number of tasks, where higher-dimensional quantum systems offer an advantage, like quantum communication 
\cite{EckerOvercoming2019,hrmo2023native,erhard2020advances} and multiparameter metrology~\cite{DuUncertaintyRelationsPRL2026}.  Moreover, it especially stands out as one of the few entanglement measures linked to quantum computational advantage \cite{VandenNestUniversal2013,VandenNestClassical2007,VidalEfficient2003}. 

Certifying the effective dimension of a system is particularly important for platforms that are formally described by an infinite-dimensional Hilbert space, which can never be fully accessed or controlled in practice due to the fundamental limitation of experimental resources (e.g. bounded energy, noise)~\cite{LvovskyContinuous2009}.
In particular, CV platforms have become one of the leading architectures for quantum technologies, from bosonic codes storing logical qubits~\cite{GottesmanKitaevPreskill2001,FluhmannEncoding2019,campagne-ibarcq_quantum_2020,SivakRealTime2023} and qudits~\cite{BrockQudits2025} in many levels of microwave-cavity and trapped-ion oscillators to optical cluster states entangling thousands of modes~\cite{AsavanantGeneration2019,LarsenDeterministic2019}. Yet, generating and precisely controlling states of effectively high dimensionality, such as those exhibiting entanglement across many Fock states, remains highly nontrivial~\cite{AdessoContinuousVariable2014,WalschaersNonGaussian2021,WeedbrookGaussian2012}.
Numerous recent experiments already realize and exploit high‑dimensional entanglement in CV degrees of freedom by discretizing spatial, temporal, or spectral modes and working within finite subspaces \cite{LibExperimental2025,HerreraValencia2020highdimensional,chang2024experimentalhighdimensional,yu2025quantum,ndagano2020imaging,BaptisteQuantifying2023,HuiGeneration2022,CourmeManipulation2023,DesignolleGenuine2021}, which further motivates robust Schmidt‑number certification in CV settings.
Criteria to characterize entanglement dimensionality in CV systems have been developed ~\cite{ShahandehOperational2013,SperlingDetermination2011}, e.g. based on the covariance matrix of quadratures; however, these are far from exhaustive and, crucially, are often outperformed in practice by truncation- and fidelity-based approaches, particularly in the non-Gaussian regime.

For a bipartite pure state $\ket{\psi}=\sum_{k=1}^{r} \sqrt{\lambda_k}\ket{u_k}\ket{v_k}$, the number $r$ of nonzero Schmidt coefficients is its Schmidt rank, i.e., the rank of the single-particle reduced density matrix. For a mixed state $\varrho$, the corresponding entanglement dimensionality is its Schmidt number~\cite{TerhalSchmidt2000,HorodeckiLewensteinBoundEntanglement},
\begin{equation}\label{eq:SNdefi}
\mathcal{SN}(\varrho):=\inf_{\mathcal{D}(\varrho)} \max_{\ket{\psi_i} \in \mathcal{D}(\varrho)} \operatorname{SR}(\ket{\psi_i}),
\end{equation}
where $\operatorname{SR}$ denotes the pure-state Schmidt rank and the infimum is taken over all pure-state decompositions $\mathcal{D}(\varrho)=\{p_i,\ket{\psi_i}\}$ with $\varrho=\sum_i p_i \ketbra{\psi_i}$. 
The Schmidt number is an entanglement monotone, that is, a nonnegative functional of the state that takes its minimum value on separable states and is non-increasing under {\it Local Operations and Classical Communication} (LOCC).
The easiest and canonical way of addressing experimental Schmidt number detection is the fidelity to a finite-dimensional target state, or, in the case of CV systems, the fidelity of a truncation of the state~\cite{FriisNatPhys19}. While this fidelity witness is theoretically simple to formulate,
it does not leverage the sophisticated measurement techniques that are specifically suited for CV systems.
Our work aims at filling this gap by deriving truncation-free entanglement dimensionality witnesses directly from phase-space correlations of CVLOOs. These criteria are accessible through standard CV measurements, and provide a more precise and robust characterization of the Schmidt number than truncation-based methods. 

At a conceptual level, the two approaches are different in what they assume about the state. A bosonic state inhabits an infinite-dimensional Hilbert space. A truncation-based certificate first selects a finite subspace and a target state inside it, so that its performance depends on this modelling choice as well as on the measured correlations. A phase-space Schmidt-number witness reverses this order. It asks how large the Schmidt number must be to reproduce the observed correlations, for a state of arbitrary support. This connects the resource theory of high-dimensional entanglement to the native toolbox of CV platforms, with potential to uncover 
phenomena specific to CV systems, e.g., states that are genuinely infinite-dimensionally entangled, yet positive under partial transposition (PPT)~\cite{HorodeckiLewensteinBoundEntanglement}. In fact, as we show, this toolbox can \space
reveal Schmidt numbers growing exponentially with the number of copies of a PPT-entangled Gaussian state, in analogy with recent finite-dimensional constructions~\cite{HuberLamiLancienMuellerHermes2018,KrebsMariPRL2024,KrebsGachechiladzePRA2026}.

{\bf Schmidt-number witnesses in DV systems.---}
First, let us recall some tools useful to characterize the entanglement dimensionality in finite dimensional systems. Typical Schmidt-number witnesses are obtained from upper bounds on the fidelity to some given (high-dimensional) target state $\ket{\Psi_T} = \sum_k \sqrt{\lambda_k} \ket{u_k}\ket{v_k}$: This idea has often been employed in experiments~\cite{FriisNatPhys19,HerreraValencia2020highdimensional,ndagano2020imaging,CourmeManipulation2023,chang2024experimentalhighdimensional,MalikHighDimensional2026,BavarescoMeasurementsInNP2018,MartinQuantifying2017}, i.e. if a preparation $\varrho$ is sufficiently close to an ideal target (say the maximally entangled state), then a lower bound $r$ on its entanglement dimensionality can be certified via $\bra{\Psi_T} \varrho \ket{\Psi_T} > \sum_{k=1}^{r-1} \lambda_k$, with $\lambda_k$ in non-increasing order.

On the other hand, it is known that not every (mixed, high-dimensional) entangled state can be witnessed via fidelities, even considering all possible pure target states~\cite{WeilenmannPRL2020BeyondFidelity}. To handle these cases, there are more sophisticated methods that have been developed~\cite{DAlessandroPRL2025,YiPRXQ2026}. Among them, as an idea relevant for the subsequent discussion we briefly introduce those based on correlations between local orthogonal observables (LOOs)~\cite{YuSixiaEntanglement2005,GittsovichUnifying2008,AsadianHeisenbergWeyl2016}. 

For a bipartite system, tensor products of two local orthogonal operator bases, one for each subsystem, form an orthonormal operator basis for the composite system.
The correlation matrix with components $[\mathcal X_\varrho]_{jk} := \av{g^{(a)}_j \otimes g^{(b)}_k}_\varrho$ encodes also information about entanglement. In fact, common witnesses of entanglement and also its dimensionality can be expressed
in terms of $\mathcal X$~\cite{FriisNatPhys19}.

For example, the so-called Computable Cross-Norm and Realignment (CCNR) criterion, which can be expressed as  $\| \mathcal{R}(\varrho)\|\leq 1$, where $\mathcal{R}$ is the realignment map and $\|\cdot\|$ denotes the trace norm,
can be also written as $\| \mathcal X \| = \sum_k \av{g_k^{(a)} \otimes g_k^{(b)}}\leq 1$ for some optimal $g_k^{(a)}$ and $g_k^{(b)}$ that bring the correlation matrix in its singular-value decomposition~\cite{YuSixiaEntanglement2005,GittsovichUnifying2008}. A similar form, but with correlations between different observables~\cite{EltschkaTothSiewertPRA2015}, can be found for the Positive Partial Transpose (PPT) criterion, which can be written as $\| \mathcal T_A(\varrho)\|\leq 1$ 
with $\mathcal T_A$
being the transposition map
applied only to the first system. The bounds on the right-hand side can then be lifted from one to $\mathcal{SN}(\varrho)$ for states with a bounded Schmidt number~\cite{EltschkaTothSiewertPRA2015,Tavakoli_2024,Liu2024bounding}.

Similar information is also contained in the cross-covariance matrix with elements $[X_\varrho]_{jk}:= [\mathcal X_\varrho]_{jk} - \av{g^{(a)}_j}_\varrho \av{g^{(b)}_k}_\varrho$, and in fact similar criteria have been found from such covariances, which are strictly more robust to noise than the CCNR criterion and can also detect PPT entangled states~\cite{GuhneCovariance2007,GittsovichUnifying2008,Liu2024bounding}.

{\bf CVLOO witnesses methods.---}
A convenient way to formulate CV entanglement witnesses is in terms of CVLOOs, i.e., LOOs labelled by a phase-space variable~\footnote{Note that there exist also discrete LOOs in CV systems, for example constructed from the Fock basis. However, in most of our discussion we focus on continuous LOOs, that fully characterize CV systems by means of quasi-probability distributions. See also End Matter and Supplemental Material~\cite{supplement}.}.
These are families of local observables $\{\mathcal{G}(\alpha)\}_{\alpha\in\mathbb{C}}$ satisfying the ortho-normalization condition
\begin{equation}\label{eq:orthonorm}
\tr\left(\mathcal{G}\left(\alpha\right) \mathcal{G}\left(\beta \right)\right) = \nnorm \   \delta^{(2)}\left(\alpha - \beta \right) ,
\end{equation}
where $\nnorm$ is the normalization.

A concrete Hermitian CVLOO can be constructed as~\footnote{Note that one could consider directly the non-hermitian displacement operators as CVLOOs and reproduce all the discussion that follows. However, following also the literature on this topic~\cite{ZhangDetectingPRL2013,qi2016nonlinear,liu2024phasespace}, we stick with the hermitian version for expressing our results in a more intuitive and simplified notation.} 
\be\label{eq:QalphaCVLOO}
\mathcal{G}(\alpha)=\frac{1+i}{2\sqrt{\pi}} D(\alpha)+\frac{1-i}{2\sqrt{\pi}} D^{\dagger}(\alpha) := \mathcal Q(\alpha) ,
\ee
where $D(\alpha) = 
e^{\alpha \hat a^\dagger - \alpha^* \hat a}$ with $\alpha \in \mathbb C$ is the displacement operator,
which defines the characteristic function $\chi_\varrho(\alpha)=\tr(D(\alpha) \varrho)$ and provides a full description of the density operator. We denoted the concrete basis in \cref{eq:QalphaCVLOO} with $\mathcal Q(\alpha)$ for clarity, to distinguish it from a generic CVLOO.
Similarly, from the definition of the Wigner
function $W_\varrho(\alpha) = \frac{2}{\pi} \tr(\varrho D(\alpha) \Pi D^\dagger(\alpha))$ where $\Pi = (-1)^{a^\dagger a}$ is the parity operator \cite{BishopDisplaced1994,MoyaCessaSeries1993,TilmaWigner2016} we see that CVLOOs can be constructed from the displaced parity.

An important feature of our approach is that we consider operators $\mathcal{Q}(\alpha)$ which admit direct experimental access in state-of-the-art CV platforms.
In systems described by a Jaynes--Cummings interaction, they can be measured by coupling the mode to a two-level ancilla and performing a qubit readout~\cite{WilkensPRA91,KimPRA98}.
With conditional displacements, both real and imaginary parts of the characteristic function can be accessed directly~\cite{asadian_probing_2014,johnson_sensing_2015,campagne-ibarcq_quantum_2020,FluhmannPRL20}.
Closely related protocols also allow for direct measurements of the Wigner function \cite{LutterbachPRL97}.

While many of these methods have been demonstrated in single-mode systems, the extension to multimode systems is straightforward. As main relevant examples, CV versions of the CCNR criterion~\cite{ZhangDetectingPRL2013,qi2016nonlinear,liu2024phasespace}, and the PPT criterion \cite{liu2024phasespace} can be found, whose separable-state bounds can be written in the form~\cite{ZhangDetectingPRL2013,qi2016nonlinear}
\begin{equation}
\label{eq:CCNR_PPT_CVLOO}
\begin{aligned}
\int_{\mathbb{C}}\langle \mathcal{G}_A(\alpha) \otimes \tilde{\mathcal{G}}_B(\alpha)\rangle \ \mathrm{d}^2 \alpha 
&\leqslant  \max \{ \|\mathcal{R}(\varrho)\| , \left\|\mathcal{T}_A(\varrho)\right\| \}  \leqslant 1 ,\end{aligned}
\end{equation}
where $\mathcal{G}_A(\alpha)$ and $\tilde{\mathcal{G}}_B(\alpha)$ are CVLOOs with a normalization satisfying $s \operatorname{tr}(\mathcal{G}_A(\alpha) \mathcal{G}_A(\beta))=(2-s) \operatorname{tr}(\tilde{\mathcal{G}}_B(\alpha) \tilde{\mathcal{G}}_B(\beta))=\delta^{(2)}(\alpha-\beta)$ with $0<s<2$~\cite{liu2024phasespace}. The parameter $s\equiv2\cos^2\theta$ sets the relative phase-space scale of the two local bases (see End Matter). The balanced value $s=1$ is optimal for exchange-symmetric states, which we consider below. A value $s\neq1$ matches the different local scales of asymmetric states, which arise, for example, from different losses or thermal noise on the two modes (see SM~\cite{supplement}).
Analogously to the DV case, for obtaining a witness equivalent to the CCNR criterion one should pick the two optimal CVLOOs 
that arise from the operator-Schmidt decomposition of the density operator~\footnote{In particular, with unit normalization as usual}. However they generally are not easily describable in phase-space~\footnote{An exception being for symmetric Gaussian states, for which choosing the CVLOOs considered in Refs.~\cite{ZhangDetectingPRL2013,qi2016nonlinear,liu2024phasespace}, which we also consider here, make \cref{eq:CCNR_PPT_CVLOO} a witness equivalent to the CCNR \cite{ZhangDetectingPRL2013}}, which is the reason why in Refs.~\cite{ZhangDetectingPRL2013,qi2016nonlinear,liu2024phasespace} the authors considered operators $\mathcal{Q}(f(\alpha))$ as in \cref{eq:QalphaCVLOO}, with an additional relabelling function $f(\alpha)$ such that $\det J_f=\pm 1$. Here $[J_f]_{kl}=\partial[f(\alpha)]_k / \partial {\alpha_l}$ with $\alpha_1 = \Re \alpha$, $\alpha_2 = \Im \alpha$ is the Jacobian of the relabelling function. This condition accommodates symplectic transformations and reflections on single modes.

These complement quadrature-based witnesses in the CV setting~\cite{SimonPeresHorodecki2000,DuanInseparability2000,WernerWolfPRL2001,hyllus2006optimal}, which are particularly suited for Gaussian states and have been used to quantify entanglement via monotones~\cite{GiedkeetAlPRL2003,MarianBures2003,MarianPRLEOF}, including the Schmidt number~\cite{ShahandehOperational2013,SperlingDetermination2011} (see also SM~\cite{supplement}).
Importantly, no witnesses based on CVLOOs, or phase-space quasi-probabilities, are so far available for certifying the Schmidt number, which is what we develop in the following. 

\begin{figure*}[t]
\centering
\includegraphics[width=0.95\linewidth]{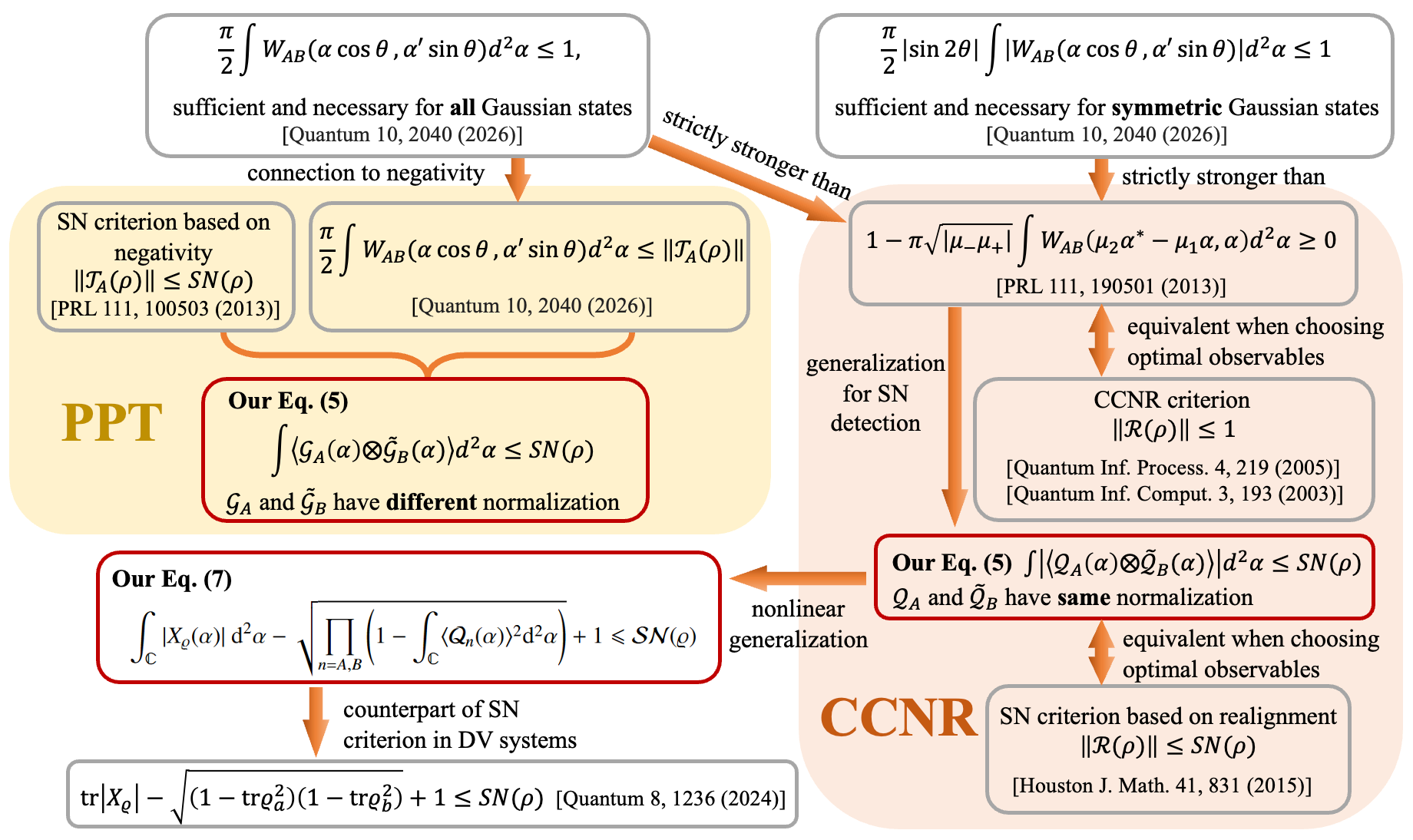}
\caption{Conceptual map illustrating the relationships between the criteria introduced in this work (red boxes) and criteria from the literature (gray boxes).}
\label{fig:ComparisonFidelityTMSTSym_SM}
\end{figure*}

{\bf Phase-space Schmidt-number criteria.---}
Here we extend witnesses based on CVLOOs to certify the Schmidt number, thereby also improving the detection of entanglement dimensionality in CV systems as compared to truncating their Hilbert space at some finite dimension.
A first way to extend \cref{eq:CCNR_PPT_CVLOO} to detect the Schmidt number is to use the inequality $\max \{\| \mathcal{R}(\varrho)\| , \| \mathcal T_A(\varrho)\|\} \leq \mathcal{SN}(\varrho)$ ~\cite{JohnstonKribs15,EltschkaNegativity2013,liu2023characterizing}. 
In this way, we provide our first Schmidt number witness, expressed in terms of phase-space correlations,
\begin{equation}
\label{eq:CCNR_PPT_CVLOO_r}
\int_{\mathbb{C}}\langle \mathcal{G}_A(\alpha) \otimes \tilde{\mathcal{G}}_B(\alpha)\rangle \ \mathrm{d}^2 \alpha \leqslant \mathcal{SN}(\varrho) ,
\end{equation}
where, as in \cref{eq:CCNR_PPT_CVLOO}, $\mathcal{G}_A(\alpha)$ and $\tilde{\mathcal{G}}_B(\alpha)$ are local CV bases, potentially with a normalization different from one. See End Matter for the details.
In what follows we derive another condition that is strictly stronger than a bound from the realigned density matrix and can detect a different set of states than the bounds in \cref{eq:CCNR_PPT_CVLOO_r}. 

As before, we consider for simplicity a bipartite system with one mode in each party, but our results are easily extended to multi-mode systems, as we also show in the SM~\cite{supplement}. 
Now we consider cross-covariances defined as
~\footnote{We omit the dependence of $X_\varrho(\alpha)$ on the concrete basis to simplify the notation.}:
\be\label{eq:Xdef}
X_\varrho(\alpha) := \av{\mathcal Q_A \otimes \mathcal Q_B}_\varrho - \av{\mathcal{Q}_A}_\varrho \av{\mathcal{Q}_B}_\varrho ,
\ee
where for concreteness we take for both parties 
$\mathcal Q_j:=\sqrt{|\det J_{f_j}|} \, \mathcal Q(f_j(\alpha))$ with $j=A,B$,
obtained from \cref{eq:QalphaCVLOO} after applying additional phase-space transformations such that $\det J_{f_A} \det J_{f_B}=\pm 1$. 
The idea behind this choice is to consider only a two-dimensional slice of the full four-dimensional phase-space, parametrized by a given $\alpha$. Intuitively, in this respect the relabelling $\alpha \mapsto f(\alpha)$ allows for having more freedom to find operators that are more aligned with the operator Schmidt bases of the density operator, while maintaining a clear phase-space description and the same experimental accessibility. Note also that such a relabelling is done locally on each mode, and therefore does not change the entanglement being probed.
See also SM~\cite{supplement}, where we discuss how these quantities can be evaluated by measuring the bipartite characteristic function along a two-dimensional plane. 

Based on these quantities, we obtain
another covariance-based witness for the Schmidt number:
\begin{equation}
\label{eq:CVSNCriterion}
\int_{\mathbb{C}} | X_\varrho(\alpha) | \  \mathrm{d}^{2} \alpha-\sqrt{\prod_{n=A,B}\left(1-\int_{\mathbb{C}}\langle \mathcal{Q}_n(\alpha)\rangle^2\mathrm{d}^{2} \alpha\right)}+1 \leqslant \mathcal{SN}(\varrho) .
\end{equation}
This condition has some similarities with \cref{eq:CCNR_PPT_CVLOO_r}, while containing additional nonlinear terms that in some cases improve over the CCNR criterion~\cite{qi2016nonlinear}.
In particular
taking $\mathcal Q_B(\alpha) = \mathcal Q_A(-\alpha^*)$ from the hermitian part of the displacement operators, the left-hand side of \cref{eq:CCNR_PPT_CVLOO_r} becomes the fidelity with respect to a CV operator 
with correlations given by 
$\av{\mathcal{Q}(\alpha) \otimes \mathcal{Q}(\alpha^\prime)}_{EPR} =\delta^{(2)}\left(\alpha^\prime + \alpha^*\right)$, which represents the limiting case of an EPR state, i.e., it is the overlap $\bra{\rm EPR}\varrho\ket{\rm EPR}$ with $\ket{\rm EPR}=\sum_n\ket{nn}$ (see SM~\cite{supplement}). This overlap is the $d\to\infty$ limit of the canonical fidelity witness $d\,\bra{\psi_+^d}\varrho\ket{\psi_+^d}\leqslant\mathcal{SN}(\varrho)$ for the maximally entangled state $\ket{\psi_+^d}=\frac{1}{\sqrt d}\sum_{n=0}^{d-1}\ket{nn}$ of the first $d$ Fock states. The nonlinear \cref{eq:CVSNCriterion}, instead, involves no target state.
For a detailed proof of these statements we refer to the End Matter. 

Intuitively, while the first term $X_\varrho(\alpha)$ accounts for correlations obtained by filtering out the local terms $\langle\mathcal Q_A(\alpha)\rangle\langle\mathcal Q_B(\alpha)\rangle$, 
the second linear-entropy term $\sqrt{\prod_{n=A,B}\bigl(1-\int_{\mathbb C}\langle\mathcal Q_n(\alpha)\rangle^2\mathrm d^2\alpha\bigr)}$ subtracts the maximal classical correlations that are allowed from the reduced-state purity $\tr(\varrho_n^2)=\int_{\mathbb C}\langle\mathcal Q_n(\alpha)\rangle^2\mathrm d^2\alpha$. These nonlinear corrections allow to improve the CCNR-based form of \cref{eq:CCNR_PPT_CVLOO_r}. 
See \cref{fig:ComparisonFidelityTMSTSym_SM} for a conceptual figure, clarifying how our new criteria are related to existing works.

The experimental requirements of our approach are also well aligned with current continuous-variable platforms. 
Importantly, our criteria do not require full two-mode tomography: it is sufficient to probe the correlations between $\mathcal Q_A(\alpha)$ and $\mathcal Q_B(f(\alpha))$ along a suitable two-dimensional slice of phase space. Testing another measure-preserving affine map $f(\alpha)=\mu\alpha+\nu\alpha^*+\gamma$ with $|\mu|^2-|\nu|^2=\pm1$ amounts to a relabelling of already sampled points or to another two-dimensional scan.
This makes the method directly compatible with state-of-the-art experiments in trapped ions \cite{leibfried_experimental_1996,johnson_sensing_2015,FluhmannPRL20,JeonPRXQ2025}, cavity and circuit QED \cite{bertet_direct_2002,vlastakis_deterministically_2013,campagne-ibarcq_quantum_2020,eickbusch_fast_2022,diringer_conditional-not_2024,valadares_-demand_2024,HeNatCommun2024}, and circuit quantum acoustodynamics \cite{satzinger_quantum_2018,von_lupke_parity_2022}, where both Gaussian and non-Gaussian entangled states are now routinely prepared and characterized. 
For optical fields, this slice follows from joint homodyne statistics. We provide explicit measurement schemes and a more detailed discussion in the SM~\cite{supplement}.

{\bf Applications.---}
We now illustrate how our method characterizes the entanglement dimensionality of continuous-variable states, and how it compares with more standard finite-dimensional truncation strategies.
As a starting point let us consider a typical Gaussian state 
that often arises naturally in noisy situations, that is 
the two-mode squeezed thermal state (TMST)\begin{equation}\label{eq:TMSTSymDef}
\varrho_{\text{TMST}}(\xi,\bar{n})=U_\xi\left(\varrho(\bar{n}) \otimes \varrho(\bar{n})\right) U^{\dagger}_\xi ,
\end{equation}
where
$\varrho(\bar{n})=\sum_{n=0}^{\infty} \frac{1}{\bar{n}+1}\left(\frac{\bar{n}}{\bar{n}+1}\right)^n \ketbra{n}$
is a single-mode thermal state with average particle number $\bar n$, and
$U_\xi = e^{\,\xi (a_1^\dagger a_2^\dagger - a_1 a_2)}$
is the two-mode squeezing operator.

\begin{figure}[t]
\centering
\includegraphics[width=\linewidth]{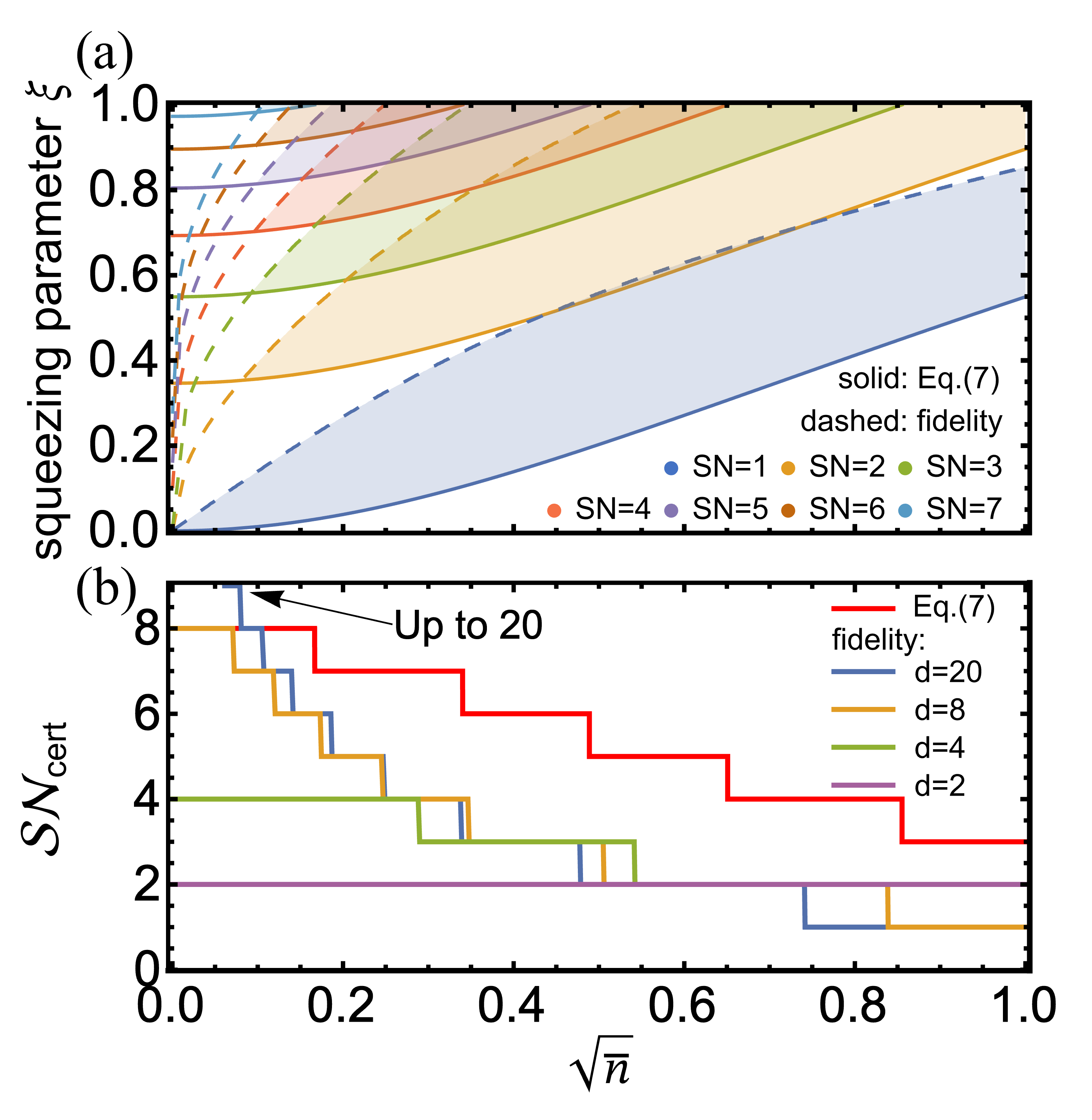}
\caption{Certified Schmidt number of the state in \cref{eq:TMSTSymDef}. (a) Upper boundaries for states with $\SN=1$ to $\SN=7$ (colour-coded) as functions of $\sqrt{\bar{n}}$ (horizontal axis) and $\xi$ (vertical axis). Solid lines correspond to the bounds from \cref{eq:CVSNCriterion}, which for this family coincide with those from \cref{eq:CCNR_PPT_CVLOO_r}; dashed lines are bounds from the fidelity to the truncated two-mode squeezed vacuum state \eqref{eq:TMSpuretruncated} with optimal truncation dimensions. Points above each curve certify a Schmidt number larger than the corresponding value. The shaded region highlights where \cref{eq:CVSNCriterion} outperforms the truncated fidelity witness. (b) Certified Schmidt number $\SN_{\rm cert}$ versus $\sqrt{\bar{n}}$ at fixed $\xi = 1$. Fidelity witnesses are shown for truncation dimensions $d = 2, 4, 8, 20$. \cref{eq:CVSNCriterion} strictly outperforms the fidelity witness even at $d = 8$. For very high truncation ($d = 20$), the fidelity witness shows slight advantage only at low thermal noise.
}
\label{fig:ComparisonFidelityTMSTSym}
\end{figure}

In the pure-state limit this reduces to the standard two-mode squeezed vacuum state
$\ket{\psi_\xi}=\sum_{n=0}^\infty \sqrt{\tfrac{(\tanh \xi)^{2n}}{(\cosh \xi)^2}\,
}\ket{nn}$, which has an infinite Schmidt number and 
whose Schmidt coefficients $\lambda_n = \frac{(\tanh \xi)^{2n}}{(\cosh \xi)^2}$ are geometrically distributed in the Fock basis. 
Note that for symmetric Gaussian states the entanglement of formation is known exactly~\cite{GiedkeetAlPRL2003,MarianPRLEOF}, as well as other entanglement measures like the Bures distance~\cite{MarianBures2003} which gives an additional benchmark for comparison.
For the TMST family, choosing $\mathcal Q_A(\alpha)$ and $\mathcal Q_B(-\alpha^*)$ as in \cref{eq:QalphaCVLOO}
maximizes the detection with \cref{eq:CCNR_PPT_CVLOO_r} and \cref{eq:CVSNCriterion} and makes the two criteria equivalent.
The certified Schmidt number with our method is then $\lceil e^{2\xi}/(2\bar n+1)\rceil$.
It is shown in \cref{fig:ComparisonFidelityTMSTSym} and, as we show in the SM~\cite{supplement}, it is larger than the lower bound inferred from the entanglement of formation~\cite{GiedkeetAlPRL2003}.

As a more conventional alternative, one may instead truncate each local Hilbert space to the first $d$ Fock states and estimate the entanglement dimensionality through the fidelity with the truncated two-mode squeezed state
\be\label{eq:TMSpuretruncated}
\ket{\psi_\xi^{(d)}} = \sum_{n=0}^{d-1} \sqrt{\mathcal N}\,\frac{(\tanh \xi)^{n}}{\cosh \xi}\ket{nn},
\ee
where $\mathcal N$ is the normalization factor.

The comparison is shown in \cref{fig:ComparisonFidelityTMSTSym}.
One can see that the truncated-fidelity witness performs well only when the state remains very close to the ideal pure target, while its performance deteriorates rapidly as thermal noise is added.
By contrast, our criterion remains effective well beyond this regime and already becomes advantageous under a slight deviation from purity. The advantage of the fidelity witness near $\bar n=0$ thus stems from its pure target.
This illustrates the main limitation of finite-dimensional truncation methods in the CV setting: they are intrinsically tied to the choice of a pure truncated target state, and therefore lose efficiency as soon as the experimentally relevant state acquires even moderate mixedness.

This behavior is not specific to Gaussian states.
To emphasize this point, we now consider an opposite limiting case, namely a flat Schmidt spectrum
$\lambda_n=1/d$ for $0\leq n\leq d-1$,
which corresponds to a finite-dimensional maximally entangled state in Fock basis
$\ket{\psi_+^d}$. Even in this case, where the fidelity witness with respect to the ideal target state is expected to be close to optimal, our method performs better, and the advantage becomes more visible when the noise is asymmetric.
Specifically, for the noisy state
\begin{equation}\label{eq:MESAsymmetricNoise}
\varrho(p,\bar n)=p\ketbra{\psi_{+}^d}+(1-p)\bigl[\varrho(\bar{n}) \otimes \varrho(0)\bigr],
\end{equation}
we find that \cref{eq:CVSNCriterion} is never worse than the fidelity witness with respect to $\ket{\psi_{+}^d}$, and can outperform it.
This is illustrated in the SM~\cite{supplement}, where we consider the case $d=5$ and $\bar n=0.5$, including the corresponding renormalization of the noise under truncation.

In both examples we use the simple choice
$\mathcal Q_B(\alpha)=\mathcal Q_A(-\alpha^*)$,
with $\mathcal Q(\alpha)$ given in \cref{eq:QalphaCVLOO} as before.
Further details on the Gaussian evaluation and on the asymmetric-noise example are given in the SM~\cite{supplement}.

For asymmetric states, it is instead essential to consider an unbalanced normalization. 
For example, for a two-mode squeezed thermal state with $\xi=1$, $\bar n_A=0$ and $\bar n_B=3$, \cref{eq:CCNR_PPT_CVLOO_r} with $s=1$ certifies $\SN\geq2$ and the optimal $s\simeq1.20$ certifies $\SN\geq5$.

Our criteria also certify high-dimensional bound entanglement. For $n$ copies of the $2\times 2$-mode Gaussian PPT-entangled state of Refs.~\cite{WernerWolfPRL2001,ZhangDetectingPRL2013}, which are PPT and therefore undistillable~\cite{HorodeckiLewensteinBoundEntanglement}, the multimode form of \cref{eq:CVSNCriterion} on a linear phase-space slice certifies $\SN\geq\lceil v^n\rceil$, with a base $1<v<2$ arbitrarily close to $2$. As PPT states, they escape any fidelity witness, with or without truncation (see SM~\cite{supplement}).

\vspace{2mm}
{\bf Conclusions and outlook.---}
In this work, we have extended the approach of quantifying entanglement through continuous-variable local orthogonal observables by focusing on the entanglement dimensionality.
This is one of the most relevant benchmarks for assessing the strength of entanglement in several applications, ranging from quantum communication and cryptography \cite{BullaNonlocal2023,BullaDistribution2023}, to quantum computing \cite{hrmo2023native}, and more foundational uses of entanglement \cite{erhard2020advances}.

From the point of view of the state preparation, bipartite CV entangled states have been experimentally demonstrated in a variety of systems.
Two-mode squeezed states have been extensively explored in optics \cite{andrews_squeezed_2015} and, more recently, also for the motional states of trapped ions \cite{MetznerPRA24}, microwave fields \cite{EichlerPRL11,FlurinPRL12} and mechanical oscillators \cite{CheungArXiv2016}.
Beyond Gaussian states, recent developments in the realization of universal gates and nonlinear dynamics across different platforms have allowed for the demonstration of non-Gaussian entanglement, including milestone optical experiments on the preparation of high-dimensional entangled states \cite{krenn2014generation,dada2011experimental,schaeff2015experimental,MartinQuantifying2017,malik2016multi,HuEfficient2020,huang2025integrated,mahmudlu2023fully,RaymondTunable2024,erhard2018experimental,wang2018multidimensional}, as well as circuit QED preparation of
entangled cats \cite{ChenScience2016}, coherent and Fock \cite{gao_entanglement_2019} states. 
Our criteria apply directly to these systems, leveraging platform-native phase-space measurements.

In typical CV experiments, especially photonic implementations, entanglement dimensionality is usually certified by truncating the Hilbert space and applying discrete-variable tools. 
Here we have shown that a genuinely continuous-variable approach gives a more effective characterization. In particular, our criteria yield a better visibility of the Schmidt number and are more robust to noise than fidelity witnesses based on finite-dimensional truncations. Our criteria also certify high-dimensional PPT entanglement that fidelity witnesses based on pure target states cannot detect.

At the same time, our results substantially strengthen phase-space and quasi-probability methods for entanglement characterization. Even for symmetric Gaussian states, the Schmidt-number lower bound obtained from our criterion can be strictly stronger than that inferred from the exact entanglement of formation. This makes our method more informative and more naturally adapted to realistic experimental conditions, and readily applicable across relevant platforms.

Our work also opens several further questions. One is whether Schmidt numbers as large as those certified here for copies of a Gaussian bound-entangled state can be certified in few-mode PPT states, Gaussian or non-Gaussian~\cite{HorodeckiLewensteinBoundEntanglement}. Another is whether analogous phase-space criteria can characterize multipartite entanglement dimensionality. More broadly, our results highlight phase-space methods as a promising route to finer quantitative characterization of CV quantum correlations.

\vspace{2mm}
{\bf Acknowledgements.---} This work is supported by the National Natural Science Foundation of China (Grants No. 12125402, No. 12534016, No. 12405005, No. 12505010, No. 12447157), Beijing Natural Science Foundation (Grant No. Z240007), the Innovation Program for Quantum Science and Technology (No. 2024ZD0302401), and the Austrian Science Fund (FWF) [\href{https://doi.org/10.55776/P35810}{10.55776/P35810}] and [\href{https://doi.org/10.55776/P36633}{10.55776/P36633}]. SL acknowledges the China Postdoctoral Science Foundation (No. 2023M740119). 
JG acknowledges the Postdoctoral Fellowship Program of CPSF (GZB20240027), and the China Postdoctoral Science Foundation (No. 2024M760072).
MF was supported by the Swiss National Science Foundation Ambizione Grant No. 208886, and The Branco Weiss Fellowship -- Society in Science, administered by the ETH Z\"{u}rich. 
GV acknowledges support from the grant No. RYC2024-048278-I funded by MCIU/AEI/10.13039/501100011033 and FSE+.
MH acknowledges funding from the Horizon-Europe research and innovation programme under grant agreement No 101070168 (HyperSpace) and the FFG through the Quantum Austria projects VANESSA-QC and MUSIQ.

\vspace{2mm}
{\bf End Matter.---}
We derive the first lower bound in \cref{eq:CCNR_PPT_CVLOO} as follows. For the complementary phase-space rescalings used here $s \operatorname{tr}(\mathcal{G}(\alpha) \mathcal{G}(\beta))=(2-s) \operatorname{tr}(\tilde{\mathcal{G}}(\alpha) \tilde{\mathcal{G}}(\beta))=\delta^{(2)}(\alpha-\beta)$ with $0<s<2$, we construct operators $\mathcal{G}^{(1)}(\alpha)=\mathcal{G}(\alpha/\sqrt{s})$ and $\tilde{\mathcal{G}}^{(1)}(\alpha)=\tilde{\mathcal{G}}(\alpha/\sqrt{2-s})$ that are normalized to $1$. Introducing the angle parameter $\theta=\arccos \sqrt{s/2}$, we obtain
\begin{equation}\label{eq:boundPPTEM}
\int_{\mathbb{C}}\langle\mathcal{G}(\alpha) \otimes \tilde{\mathcal{G}}(\alpha)\rangle \mathrm{d}^2 \alpha =\frac{1}{2} \int_{\mathbb{C}}\left\langle\mathcal{G}^{(1)}(\alpha \cos \theta) \otimes \tilde{\mathcal{G}}^{(1)}(\alpha \sin \theta)\right\rangle \mathrm{d}^2 \alpha
\end{equation}
Thus $s=2\cos^2\theta$ and $2-s=2\sin^2\theta$ rescale the arguments of $\mathcal{G}^{(1)}$ and $\tilde{\mathcal{G}}^{(1)}$ to $\alpha\cos\theta$ and $\alpha\sin\theta$, respectively. The balanced value $s=1$ ($\theta=\pi/4$) treats the two parties symmetrically, and $s\neq1$ changes their relative phase-space scale, which can be advantageous for asymmetric states.
When both $\mathcal{G}$ and $\tilde{\mathcal{G}}$ are constructed from displaced-parity operators $\frac{2}{\pi} D(\alpha) \Pi D^\dagger(\alpha)$, the proof that the expression in \cref{eq:boundPPTEM} gives a lower bound to $\|\mathcal{T}_A(\varrho)\|$  
has been established in \cite{liu2024phasespace}. 
For the choice $\tilde{\mathcal{G}}^{(1)}(\alpha)=\mathcal{Q}_B(-\alpha^*)$ used here, the SM gives an explicit proof: the left-hand side of \cref{eq:boundPPTEM} is the expectation value of the parity of a beam-splitter mode in the partially transposed operator.

Here we point out some short details on the transformations between various LOOs in CV systems. Given a discrete basis such as the Fock basis we can construct the operators $g_{\mu} = \ketbra{n}{m}$ with $\mu=(nm)$ being a compact label for both indices. These operators form an orthonormal basis, as they satisfy $\tr(g_\mu^\dagger g_\nu) = \delta_{\mu \nu}$. 
A change of basis, e.g., from a discrete to a continuous one, is done via an orthogonal kernel $\mathcal{Q}(\alpha)=\sum_\mu O_\mu(\alpha) g_\mu$ with 
$O_\mu(\alpha) = \tr(\mathcal{Q}(\alpha) g_\mu^\dagger)$. This transformation satisfies $\sum_\mu O^\dagger_\mu(\alpha) O_\mu(\beta)= \delta^{(2)}(\alpha -\beta)$
and also $\intdalpha{O^\dagger_\mu(\alpha) O_\nu(\alpha)} = \delta_{\mu \nu}$.
Analogous transformations allow to change between two continuous or two discrete bases.

Now, we can give the explicit proof of the criterion in \cref{eq:CVSNCriterion}. This is based on the following Lemma, 
that provides a bound on the absolute value of the cross-covariances for pure states with a given Schmidt number.

\begin{lemma}\label{lemma:1}
Given any two CVLOOs $\mathcal Q_A(\alpha)$ and $\mathcal Q_B(\alpha)$, for pure bipartite states $\ket{\psi_r}$ with Schmidt rank not larger than $r$, the cross-covariances satisfy
\begin{equation}\label{eq:lemma1}
\intdalpha{\left|\Gamma_{\psi_r} (\mathcal Q_A (\alpha),\mathcal Q_B(\alpha))\right|} \leqslant r-\tr\left(\left[\tr_B \ketbra{\psi_r}\right]^2\right) .
\end{equation}
\end{lemma}

{\it Proof.---}Let us consider a pure Schmidt-rank-$r$ state in its Schmidt decomposition: $\ket{\psi_r}=\sum_{n=0}^{r-1} \sqrt{\lambda_n} \ket{n n}$, and let us take for each mode the discrete operator basis $g_{\mu=(nm)} = \ketbra{n}{m}$. 
In this case, the nonzero cross-covariances of those operators form a (finite-dimensional) $r^2 \times r^2$ matrix 
$\Gamma_{\psi_r} (g_\mu^{(a)} , g_\nu^{(b)}) := (X_{\psi_r})_{\mu \nu} = \av{(g^{(a)}_\mu)^\dagger \otimes g^{(b)}_\nu} - \av{g^{(a)}_\mu}\av{g^{(b)}_\nu}$, 
where $\mu = (n,m)$ is a double-index with $0\leq n , m \leq (r-1)$. Moreover, the following bound holds: $\| X_{\psi_r} \| = \sum_{\mu=1}^{r^2} \Lambda_\mu = \sum_{n,m=0}^{r-1} \sqrt{\lambda_n \lambda_m} - \sum_n \lambda^2_n \leq r  - \sum_n \lambda^2_n = r-\tr\left(\left[\tr_B \ketbra{\psi_r}\right]^2\right)$, where we indicated with $\Lambda_\mu$ the singular values of the matrix $X_{\psi_r}$. This bound is essentially due to $\sum_{nm} \sqrt{\lambda_n \lambda_m} \leq r$, plus the fact that the submatrix $D_{nm} := \lambda_n\delta_{nm} - \lambda_{n}\lambda_m$ is positive semidefinite~\cite{Liu2024bounding}. 
Now, in order to obtain \cref{eq:lemma1} we have to make a change of CVLOOs from the (discrete) one that gives the singular value decomposition of $X_{\psi_r}$, which we indicate as $\tilde g_\mu$, to the (continuous) ones $\mathcal{Q}(\alpha)$.
Applying the transformation described earlier we get 
$\intdalpha{\left|\Gamma_{\psi_r} (\mathcal Q_A (\alpha),\mathcal Q_B(\alpha))\right|}  = \intdalpha{\left|\sum_{\mu} \tr[\mathcal{Q}_A(\alpha) \tilde g_\mu^{(a)}] \tr[\mathcal{Q}_B(\alpha) (\tilde g^{(b)}_\mu)^\dagger] \Lambda_\mu \right|} \leq \sum_\mu \Lambda_\mu = \| X_{\psi_r} \|$, where we have used the triangle inequality and the Cauchy–Schwarz inequality, exploiting the orthogonality of the basis changes.
\qed

The CCNR criterion can be also expressed via correlations in a CVLOO. In particular, a similar reasoning as in the above proof can be made considering the cross-correlations $(C_{\psi_r})_{\mu \nu} = \av{(g^{(a)}_\mu)^\dagger \otimes g^{(b)}_\nu}$, 
that for a Schmidt-rank-$r$ pure state also form a $r^2 \times r^2$ matrix with a bounded trace norm
\be
\| C_{\psi_r} \| \leq r ,
\ee
which gives a relation that can be similarly extended to all CVLOOs due to the orthogonality of the basis change. This condition is nothing but $\| \mathcal R(\varrho) \| \leq r$~\cite{JohnstonKribs15,ZhangDetectingPRL2013}. 
One can then use the simple argument $\intdalpha{\av{\mathcal Q_A(\alpha) \otimes \mathcal Q_B(\alpha)}_{\psi_r}} \leq \| C_{\psi_r} \| \leq r$ valid for pure states and extend it to mixed states due to the linearity of the expression.
In this way, one can then find a Schmidt-number condition as in \cref{eq:CCNR_PPT_CVLOO_r} coming from the CCNR.
A related argument can be found to bound the norm of the partially transposed matrix and obtain again an expression as in \cref{eq:CCNR_PPT_CVLOO_r}, this time with differently normalized 
observables~\cite{liu2024phasespace}, as we have also discussed earlier.

Based on \cref{lemma:1} the proof of \cref{eq:CVSNCriterion} follows. Once again, in this work we focus on specific CVLOOs, but 
the proof works in a more general scenario of two arbitrary bases for the two parties.

\textit{Proof of \cref{eq:CVSNCriterion}---}Let us consider a mixed bipartite state decomposed as $\varrho= \sum_k p_k \ketbra{\psi_k}$ where all the $\ket{\psi_k}$ have Schmidt rank smaller than or equal to a given $r$. 
Let us consider the functions
\be
\begin{aligned}
\mathcal{A}_\varrho(\alpha) &= \var{(\mathcal{Q}_A(\alpha)\otimes \id)}_\varrho - \sum_k p_k \var{(\mathcal{Q}_A(\alpha) \otimes \id)}_{\psi_k} , \\
\mathcal{B}_\varrho(\alpha) &= \var{(\id \otimes \mathcal{Q}_B(\alpha))}_\varrho - \sum_k p_k \var{(\id \otimes \mathcal{Q}_B(\alpha))}_{\psi_k} , \\
\mathcal{C}_\varrho(\alpha) &= \Gamma_{\varrho} (\mathcal Q_A (\alpha),\mathcal Q_B(\alpha)) - \sum_k p_k \Gamma_{\psi_k} (\mathcal Q_A (\alpha),\mathcal Q_B(\alpha)) .
\end{aligned}
\ee
Now, we have that the inequality
\be
\mathcal{A}_\varrho(\alpha) \mathcal{B}_\varrho(\alpha) \geq |\mathcal{C}_\varrho(\alpha)|^2
\ee
follows from the positivity of the matrix 
\be
\Delta := 
\begin{pmatrix}
\mathcal{A}_\varrho(\alpha) & \mathcal{C}_\varrho(\alpha) \\
\mathcal{C}_\varrho(\alpha) & \mathcal{B}_\varrho(\alpha)
\end{pmatrix}  \geq 0 ,
\ee
which is in turn a consequence of the concavity and the positivity of covariance matrices.

This also implies that the following function is positive for all $\vec \alpha \in \mathbb C$ and $t \in \mathbb{R}$:
\be\label{eq:Tdef1}
T_\varrho(\alpha, t) := \mathcal{A}_\varrho(\alpha) + 4t^2 \mathcal{B}_\varrho(\alpha) - 4 |t| |\mathcal{C}_\varrho(\alpha)| \geq 0 ,
\ee
and so is its integral over $\alpha$: $\int_{\mathbb C} T_\varrho(\alpha, t) \ \mathrm{d}^2 \alpha \geq 0$. 
Now, from the fact that for single modes it holds that
$S_L(\varrho) := 1-\tr(\varrho^2)= 1-\int_{\mathbb{C}} \langle \mathcal{Q}(\alpha) \rangle_\varrho^2 \mathrm{d}^2 \alpha$, we can find the following relations
\be
\begin{aligned}
\intdalpha{\mathcal{A}_\varrho(\alpha)} &= S_L(\varrho_A) - \sum_k p_k S_L(\tr_B(\ketbra{\psi_k}{\psi_k}))  , \\
\intdalpha{\mathcal{B}_\varrho(\alpha)} &= S_L(\varrho_B) - \sum_k p_k S_L(\tr_B(\ketbra{\psi_k}{\psi_k})) , 
\end{aligned}
\ee
and moreover we have that $\intdalpha{|\mathcal{C}_\varrho(\alpha)|} \geq \intdalpha{\left(|X_\varrho(\alpha)|- \sum_k p_k |X_{\psi_k}(\alpha)|\right)}$, where now
we introduced the notation $\Gamma_{\varrho} (\mathcal Q_A (\alpha),\mathcal Q_B(\alpha)):= X_\varrho(\alpha)$ for simplicity.
This last expression can be in turn bounded by using the inequality \cref{eq:lemma1} from \cref{lemma:1} as
\be
\intdalpha{|\mathcal{C}_\varrho(\vec \alpha)|} \geq \intdalpha{|X_{\varrho} (\alpha)|} - r + \sum_k p_k \tr\left(\left[\tr_B\left|\psi_k\right\rangle\left\langle\psi_k\right|\right]^2\right) .
\ee
Substituting these relations into \cref{eq:Tdef1} and using $S_L(\varrho)\geq 0$, we find
\be
\tilde{T}_\varrho (t) := S_L(\varrho_A) + 4t^2 S_L(\varrho_B) -4 |t| \left(\intdalpha{|X_\varrho(\alpha)|} - r +1\right) \geq 0 ,
\ee
which holds for all real values of $t$.
Next, minimizing this function over all $t\in \mathbb{R}$ we find 
\be
\intdalpha{|X_\varrho(\alpha)|} - \sqrt{S_L(\varrho_A) S_L(\varrho_B)} +1 \leq r ,
\ee
which results from the fact that the minimum of $\tilde{T}_\varrho(t)$ is achieved for $2|t|=\left(\intdalpha{|X_\varrho(\alpha)|} - r +1\right)/S_L(\varrho_B)$ when $\intdalpha{|X_\varrho(\alpha)|} - r +1>0$. If this quantity is nonpositive, the desired inequality holds trivially. \qed

\bibliography{references}

\appendix
\begin{widetext}
\newpage
\clearpage
\def\thefigure{S\arabic{figure}}
\setcounter{figure}{0}
\renewcommand{\theequation}{S\arabic{equation}}
\setcounter{equation}{0}
\renewcommand{\thetable}{S\arabic{table}}
\setcounter{table}{0}
\renewcommand{\theproposition}{S\arabic{proposition}}
\setcounter{proposition}{0}
\makeatletter
\@ifundefined{theHequation}{}{\renewcommand{\theHequation}{SM.\arabic{equation}}}
\@ifundefined{theHfigure}{}{\renewcommand{\theHfigure}{SM.\arabic{figure}}}
\@ifundefined{theHtable}{}{\renewcommand{\theHtable}{SM.\arabic{table}}}
\@ifundefined{theHproposition}{}{\renewcommand{\theHproposition}{SM.\arabic{proposition}}}
\makeatother

\section*{Supplemental Material}

This Supplemental Material (SM) collects the background and the derivations behind the statements of the main text. Section~I fixes the notation: phase-space representations, the continuous-variable local orthogonal observables (CVLOOs) used throughout, and the standard form of Gaussian states under local operations. Section~II discusses general properties of the two Schmidt-number criteria of the main text, the linear \cref{eq:CCNR_PPT_CVLOO_r} and the nonlinear \cref{eq:CVSNCriterion}: their multimode form, the interpretation of the linear criterion as a parity measurement on the partially transposed state, which also clarifies the role of the normalization parameter $s$, and the relation between the two criteria. Sections~III and~IV apply the criteria to Gaussian and to non-Gaussian states, respectively, and compare them with other methods. Section~V shows that the criteria certify Schmidt numbers that grow exponentially with the number of copies of a Gaussian bound-entangled state. Section~VI discusses experimental implementations.

For orientation, we briefly recall the criteria from the literature shown in \cref{fig:ComparisonFidelityTMSTSym_SM}. The most common entanglement tests for continuous-variable (CV) systems use second moments of the quadratures, most notably the positive-partial-transpose (PPT) criterion for covariance matrices~\cite{SimonPeresHorodecki2000,DuanInseparability2000}, which is necessary and sufficient for two-mode Gaussian states~\cite{SimonPeresHorodecki2000,WernerWolfPRL2001}, and its generalizations to other combinations of quadratures~\cite{GiovannettiManciniVitaliTombesi2003,hyllus2006optimal}. For Gaussian states these tools also give entanglement measures, such as the entanglement of formation of symmetric states~\cite{GiedkeetAlPRL2003,MarianPRLEOF} and the Bures distance~\cite{MarianBures2003}, as well as a hierarchy of Schmidt-number witnesses built from quadratic operators, whose thresholds have to be computed numerically for each case~\cite{ShahandehOperational2013}. Other Schmidt-number criteria are fidelities with specific target states, e.g., with Fock-basis Schmidt bases~\cite{SperlingDetermination2011}. Phase-space quasi-probabilities give instead entanglement witnesses built from CVLOOs~\cite{ZhangDetectingPRL2013,qi2016nonlinear,liu2024phasespace}, which after optimization are equivalent to CV versions of the realignment and PPT criteria, detect Gaussian bound-entangled states~\cite{ZhangDetectingPRL2013}, and bound the negativity~\cite{liu2024phasespace}, its convex-roof extension, and the entanglement of formation~\cite{ZhangDetectingPRL2013}.
The criteria of the main text extend this last approach to the Schmidt number.

\section*{I. Notation and phase-space tools}

\subsection*{A. Displacements, characteristic and Wigner functions}

For a single mode with annihilation operator $\hat a$ we use the quadratures $\hat x=\hat a+\hat a^\dagger$ and $\hat p=i(\hat a^\dagger-\hat a)$, so that $[\hat x,\hat p]=2i$ and the vacuum has unit variances. The displacement operator
\be\label{eqS:displacement}
D(\alpha)=e^{\alpha\hat a^\dagger-\alpha^*\hat a}=e^{\,i(\hat x\,\mathrm{Im}\,\alpha-\hat p\,\mathrm{Re}\,\alpha)},\qquad \alpha\in\mathbb C ,
\ee
satisfies $D^\dagger(\alpha)=D(-\alpha)$ and $\tr[D^\dagger(\alpha)D(\beta)]=\pi\,\delta^{(2)}(\alpha-\beta)$~\cite{CahillOrdered1969}. The displacements are thus a continuous operator basis, and a state $\varrho$ is fully described by its characteristic function
\be\label{eqS:chi}
\chi_\varrho(\alpha)=\tr[\varrho\,D(\alpha)] ,
\ee
which satisfies $\chi_\varrho(0)=1$, $\chi_\varrho(-\alpha)=\chi_\varrho(\alpha)^*$ and $\tr(\varrho^2)=\frac1\pi\int_{\mathbb C}|\chi_\varrho(\alpha)|^2\,\de^2\alpha$.
Equivalently, the Wigner function is the expectation value of a displaced parity operator, $W_\varrho(\alpha)=\frac2\pi\tr[\varrho\,D(\alpha)\Pi D^\dagger(\alpha)]$ with $\Pi=(-1)^{\hat a^\dagger\hat a}$~\cite{BishopDisplaced1994,MoyaCessaSeries1993,TilmaWigner2016}.
The following identity is also used repeatedly below:
\be\label{eqS:identities}
\int_{\mathbb C}D(\alpha)\,\de^2\alpha=2\pi\,\Pi ,
\ee
which expresses $W_\varrho(0)=\frac2\pi\tr(\varrho\Pi)$ through the characteristic function.

For $M$ modes, $D(\vec\alpha)=\bigotimes_{j=1}^M D(\alpha_j)$ with $\vec\alpha\in\mathbb C^M$, $\chi_\varrho(\vec\alpha)=\tr[\varrho D(\vec\alpha)]$, and $\tr[D^\dagger(\vec\alpha)D(\vec\beta)]=\pi^M\delta^{(2M)}(\vec\alpha-\vec\beta)$. We collect the quadratures in $\hat{\vec r}=(\hat x_1,\hat p_1,\dots,\hat x_M,\hat p_M)$ and the phase-space labels in the real vector $\vec y=(\mathrm{Re}\,\alpha_1,\mathrm{Im}\,\alpha_1,\dots,\mathrm{Re}\,\alpha_M,\mathrm{Im}\,\alpha_M)\in\mathbb R^{2M}$, so that $D(\vec\alpha)=\exp(i\hat{\vec r}^{\mathsf T}\Omega_M\vec y)$ and $[\hat r_j,\hat r_k]=2i\,[\Omega_M]_{jk}$, with the symplectic form $\Omega_M=\bigoplus_{j=1}^M\omega$, $\omega=\bigl(\begin{smallmatrix}0&1\\-1&0\end{smallmatrix}\bigr)$. We use $\vec\alpha$ and $\vec y$ interchangeably as arguments.

\subsection*{B. The CVLOOs used in this work}

Throughout this SM we use a single family of CVLOOs, the Hermitian combinations of displacements of main-text \cref{eq:QalphaCVLOO},
\be\label{eqS:Q}
\mathcal Q(\alpha)=\frac{1+i}{2\sqrt\pi}\,D(\alpha)+\frac{1-i}{2\sqrt\pi}\,D^\dagger(\alpha) ,
\ee
which satisfy $\tr[\mathcal Q(\alpha)\mathcal Q(\beta)]=\delta^{(2)}(\alpha-\beta)$ and form a complete operator basis. Their expectation values follow from the characteristic function,
\be\label{eqS:Qmean}
\av{\mathcal Q(\alpha)}_\varrho=\frac{\mathrm{Re}\,\chi_\varrho(\alpha)-\mathrm{Im}\,\chi_\varrho(\alpha)}{\sqrt\pi} ,
\ee
and the purity of a state is $\tr(\varrho^2)=\int_{\mathbb C}\av{\mathcal Q(\alpha)}_\varrho^2\,\de^2\alpha$, since the cross term $\mathrm{Re}\,\chi_\varrho\,\mathrm{Im}\,\chi_\varrho$ is odd in $\alpha$. For two parties $A$ and $B$, with $\chi_\varrho(\alpha,\beta)=\tr[\varrho\,D(\alpha)\otimes D(\beta)]$, one finds similarly
\be\label{eqS:Qcorr}
\av{\mathcal Q_A(\alpha)\otimes\mathcal Q_B(\beta)}_\varrho=\frac{\mathrm{Re}\,\chi_\varrho(\alpha,-\beta)-\mathrm{Im}\,\chi_\varrho(\alpha,\beta)}{\pi} ,
\ee
where the subscript indicates the mode on which $\mathcal Q$ acts. For states with a real and even characteristic function, such as the zero-mean Gaussian states below, these relations reduce to $\av{\mathcal Q(\alpha)}_\varrho=\chi_\varrho(\alpha)/\sqrt\pi$ and $\av{\mathcal Q_A(\alpha)\otimes\mathcal Q_B(\beta)}_\varrho=\chi_\varrho(\alpha,-\beta)/\pi$. For $M$ modes, $\mathcal Q(\vec\alpha)$ is defined as in \cref{eqS:Q} with $D(\vec\alpha)$ and with $\sqrt\pi$ replaced by $\pi^{M/2}$, and all the relations above hold with the same replacement.

Two modifications of \cref{eqS:Q} appear in the main text, and both give again CVLOOs:
\begin{itemize}
\item[(i)] \textit{Relabelled CVLOOs} $\mathcal Q(f(\alpha))$, where $f$ is a phase-space map with Jacobian determinant $|\det J_f|=1$, are orthonormal as well, since then $\delta^{(2)}(f(\alpha)-f(\beta))=\delta^{(2)}(\alpha-\beta)$. Examples are the reflection $f(\alpha)=-\alpha^*$, which we call the \textit{EPR pairing} (see Sec.~II\,B), and the affine maps $f(\alpha)=\mu\alpha+\nu\alpha^*+\gamma$ with $|\mu|^2-|\nu|^2=\pm1$. More generally, $\sqrt{|\det J_g|}\,\mathcal Q(g(\alpha))$ is a unit-normalized CVLOO family for every invertible linear map $g$. Sec.~III\,D uses such a rescaled family.
\item[(ii)] \textit{Rescaled arguments.} The CVLOOs $\mathcal Q(\sqrt s\,\alpha)$ satisfy $s\,\tr[\mathcal Q(\sqrt s\,\alpha)\mathcal Q(\sqrt s\,\beta)]=\delta^{(2)}(\alpha-\beta)$. In this work, the CVLOOs $\mathcal G_A$ and $\tilde{\mathcal G}_B$ with the unbalanced normalization of main-text \cref{eq:CCNR_PPT_CVLOO_r} are
\be\label{eqS:GfromQ}
\mathcal G_A(\alpha)=\mathcal Q_A(\sqrt s\,\alpha) ,
\qquad
\tilde{\mathcal G}_B(\alpha)=\mathcal Q_B\bigl(\sqrt{2-s}\,f(\alpha)\bigr) ,
\qquad 0<s<2 ,
\ee
with a linear relabelling $f$, and $s=1$ is the balanced case.
\end{itemize}
To avoid any ambiguity, all CVLOOs in this SM are written in terms of $\mathcal Q$; the symbols $\mathcal G$ of the main text appear only in \cref{eqS:GfromQ}, which relates them to $\mathcal Q$. Besides CVLOOs, one can also use discrete local orthonormal operator bases, such as the Fock-basis operators $\ket{n}\bra{m}$. We recall that the change of basis between the two is given in the End Matter, where it is used to prove the nonlinear criterion.

\subsection*{C. Gaussian states and their standard form under local operations}

A state is Gaussian if its characteristic function is Gaussian~\cite{WeedbrookGaussian2012},
\be\label{eqS:chiGauss}
\chi_\varrho(\vec y)=\exp\Bigl[-\tfrac12\,(\Omega_M\vec y)^{\mathsf T}\,\Gamma\,(\Omega_M\vec y)+i\,\vec d^{\mathsf T}\Omega_M\vec y\Bigr] ,
\ee
where $\vec d=\av{\hat{\vec r}}_\varrho$ is the mean vector and $\Gamma_{jk}=\frac12\av{\{\hat r_j-d_j,\hat r_k-d_k\}}_\varrho$ is the covariance matrix (CM), which satisfies $\Gamma+i\Omega_M\geq0$ and equals the identity for the vacuum. For a bipartite state we write 
\be
\Gamma=\left(\begin{matrix}A&C\\C^{\mathsf T}&B\end{matrix}\right), 
\ee
where $A$ and $B$ are the CMs of the reduced states and $C$ contains the correlations between the parties.

\textit{Standard form.} Local Gaussian unitaries $U_A\otimes U_B$ displace the mean vector and transform the CM as $\Gamma\mapsto S\Gamma S^{\mathsf T}$, with $S=S_A\oplus S_B$ and symplectic $S_A,S_B$, i.e., $S_A\Omega S_A^{\mathsf T}=\Omega$~\cite{WeedbrookGaussian2012}. They change neither the entanglement nor the Schmidt number, so the state can be brought to a convenient standard form. First, local displacements set $\vec d=0$. Second, by Williamson's theorem applied to each party, $S_A$ and $S_B$ can be chosen such that the local blocks become diagonal,
\be\label{eqS:Williamson}
A=\bigoplus_j a_j I_2 ,
\qquad
B=\bigoplus_k b_k I_2 ,
\ee
where $a_j,b_k\geq1$ are the local symplectic eigenvalues. In this form, each reduced state is a product of thermal states with $(a_j-1)/2$ and $(b_k-1)/2$ mean photons, and all the information on the correlations between the parties is contained in $C$. Local passive transformations that preserve the blocks~\eqref{eqS:Williamson} can then be used to simplify $C$. For one mode per party, a local phase rotation of each mode diagonalizes $C$, which gives the standard form~\cite{SimonPeresHorodecki2000,DuanInseparability2000}
\be\label{eqS:standardform}
\Gamma=\begin{pmatrix}
a & 0 & c_1 & 0\\
0 & a & 0 & c_2\\
c_1 & 0 & b & 0\\
0 & c_2 & 0 & b
\end{pmatrix}.
\ee
A two-mode Gaussian state with $c_1c_2\geq0$ is PPT, and hence separable~\cite{SimonPeresHorodecki2000}, which means that for entangled states we can take $c_1>0>c_2$. 

\textit{Partial transposition.} In phase space, the partial transposition of $B$ is the reflection $\hat p_B\mapsto-\hat p_B$ of the momenta of $B$, which maps $\Gamma$ to $\widetilde\Gamma=\Lambda\Gamma\Lambda$, with $\Lambda=\diag(1,1,1,-1)$ for two modes~\cite{SimonPeresHorodecki2000}. A two-mode Gaussian state is entangled if and only if $\widetilde\Gamma+i\Omega_2\not\geq0$, i.e., if and only if the smallest symplectic eigenvalue $\tilde\nu_-$ of $\widetilde\Gamma$ is smaller than one. In this case~\cite{VidalWernerComputable2002,AdessoIlluminatiPRA2005}
\be\label{eqS:negativity}
\|\mathcal T_A(\varrho)\|=\|\mathcal T_B(\varrho)\|=\frac{1}{\tilde\nu_-} ,
\ee
which enters the negativity bound on the Schmidt number, $\|\mathcal T_A(\varrho)\|\leq\SN(\varrho)$~\cite{EltschkaNegativity2013,liu2023characterizing}. For the standard form~\eqref{eqS:standardform},
\be\label{eqS:numinus}
\tilde\nu_-^2=\tfrac12\Bigl[\tilde\Delta-\sqrt{\tilde\Delta^2-4\det\Gamma}\Bigr],
\qquad \text{with} \quad 
\tilde\Delta=a^2+b^2-2c_1c_2 ,
\ee
which gives $\tilde\nu_-=\sqrt{(a-c_1)(a+c_2)}$ for symmetric states, $a=b$, and $\tilde\nu_-=\frac12\bigl[a+b-\sqrt{(a-b)^2+4c^2}\bigr]$ for $c_1=-c_2=c$.

\section*{II. General properties of the criteria}

With the CVLOOs of \cref{eqS:GfromQ}, the two criteria of the main text, \cref{eq:CCNR_PPT_CVLOO_r,eq:CVSNCriterion}, read
\begin{align}
\label{eqS:linear}
\mathcal L_s^f(\varrho)&:=\int_{\mathbb C}\av{\mathcal Q_A(\sqrt s\,\alpha)\otimes\mathcal Q_B\bigl(\sqrt{2-s}\,f(\alpha)\bigr)}_\varrho\,\de^2\alpha\;\leq\;\SN(\varrho) ,\\
\label{eqS:nonlinear}
\mathcal N^f(\varrho)&:=\int_{\mathbb C}\bigl|X_\varrho^f(\alpha)\bigr|\,\de^2\alpha-\sqrt{S_L(\varrho_A)\,S_L(\varrho_B)}+1\;\leq\;\SN(\varrho) ,
\end{align}
with the cross-covariance $X_\varrho^f(\alpha)=\av{\mathcal Q_A(\alpha)\otimes\mathcal Q_B(f(\alpha))}_\varrho-\av{\mathcal Q_A(\alpha)}_\varrho\av{\mathcal Q_B(f(\alpha))}_\varrho$ of main-text \cref{eq:Xdef} and the linear entropies $S_L(\varrho_j)=1-\tr(\varrho_j^2)=1-\int_{\mathbb C}\av{\mathcal Q_j(\alpha)}_\varrho^2\,\de^2\alpha$ of the reduced states, $j=A,B$. Since the Schmidt number is an integer, the criteria certify $\SN(\varrho)\geq\lceil\mathcal L_s^f(\varrho)\rceil$ and $\SN(\varrho)\geq\lceil\mathcal N^f(\varrho)\rceil$, respectively, which we call the certified Schmidt number $\SN_{\rm cert}$.

\subsection*{A. Multimode systems}

All definitions extend directly to $M$ modes per party, with labels $\vec\alpha\in\mathbb C^M$, the multimode CVLOOs $\mathcal Q(\vec\alpha)$ of Sec.~I\,B, and a relabelling $f$ of $\mathbb C^M\simeq\mathbb R^{2M}$ with $|\det J_f|=1$. The proof of the nonlinear criterion in the End Matter uses only the Schmidt decomposition across the bipartition, the orthonormality and completeness of the local operator bases, and the purity formula $\tr(\varrho^2)=\int\av{\mathcal Q(\vec\alpha)}_\varrho^2\,\de^{2M}\alpha$, and therefore holds unchanged:
\be\label{eqS:nonlinearM}
\mathcal N^f(\varrho)=\int_{\mathbb C^M}\bigl|X_\varrho^f(\vec\alpha)\bigr|\,\de^{2M}\alpha-\sqrt{S_L(\varrho_A)\,S_L(\varrho_B)}+1\leq\SN(\varrho) ,
\ee
and the same holds for the balanced linear criterion, $s=1$, which for unit-normalized local families is bounded by the realignment norm~\cite{ZhangDetectingPRL2013,JohnstonKribs15}. The criteria thus require the correlations only on a $2M$-dimensional slice of the $4M$-dimensional phase space. 

\subsection*{B. Parity interpretation of the linear criterion}

The linear criterion in \cref{eqS:linear} can be seen as a parity measurement on the partially transposed state, which also gives a direct proof of main-text \cref{eq:CCNR_PPT_CVLOO_r} for all $0<s<2$. Let $\cos\theta=\sqrt{s/2}$, as in the End Matter, and consider the beam-splitter mode $\hat c_\theta=\cos\theta\,\hat a-\sin\theta\,\hat b$ with parity $\Pi_{c_\theta}=(-1)^{\hat c_\theta^\dagger\hat c_\theta}$. For every state $\varrho$,
\be\label{eqS:parity}
\mathcal L_s(\varrho)=\tr\bigl[\mathcal T_B(\varrho)\,\Pi_{c_\theta}\bigr]\leq\|\mathcal T_B(\varrho)\|\leq\SN(\varrho) .
\ee
\textit{Proof.} By \cref{eqS:Qcorr}, the integrand of $\mathcal L_s$ is $\frac1\pi\bigl[\mathrm{Re}\,\chi_\varrho(\sqrt s\,\alpha,\sqrt{2-s}\,\alpha^*)-\mathrm{Im}\,\chi_\varrho(\sqrt s\,\alpha,-\sqrt{2-s}\,\alpha^*)\bigr]$. Because $\chi_\varrho(-\vec y)=\chi_\varrho(\vec y)^*$, all imaginary parts are odd under $\alpha\mapsto-\alpha$ and integrate to zero, so that $\mathcal L_s(\varrho)=\frac1\pi\int_{\mathbb C}\chi_\varrho(\sqrt s\,\alpha,\sqrt{2-s}\,\alpha^*)\,\de^2\alpha$ 
and since $D(\alpha)^{\mathsf T}=D(-\alpha^*)$ for the transposition in the Fock basis, which follows from $\hat a^{\mathsf T}=\hat a^\dagger$, we have
$\chi_\varrho(\alpha_1,\alpha_2^*)=\tr[\mathcal T_B(\varrho)\,D(\alpha_1)\otimes D(-\alpha_2)]$. Afterwards, we can see that the two-mode displacement $D(\sqrt2\cos\theta\,\alpha)\otimes D(-\sqrt2\sin\theta\,\alpha)=D_{c_\theta}(\sqrt2\,\alpha)$ displaces only the mode $\hat c_\theta$, and from \cref{eqS:identities} $\frac1\pi\int_{\mathbb C}D_{c_\theta}(\sqrt2\,\alpha)\,\de^2\alpha=\Pi_{c_\theta}$. This proves the equality in \cref{eqS:parity}. The first inequality holds because $\Pi_{c_\theta}$ has eigenvalues $\pm 1$, and the second one is the Schmidt-number bound $\|\mathcal T_B(\varrho)\|=\|\mathcal T_A(\varrho)\|\leq\SN(\varrho)$~\cite{EltschkaNegativity2013,liu2023characterizing}. \qed

This proves main-text \cref{eq:CCNR_PPT_CVLOO_r} for the CVLOOs~\eqref{eqS:GfromQ} with $s=1$ and $f(\alpha) = -\alpha^*$. The same proof can be then extended to the case $s\neq1$, by using the relation between the two local families in \cref{eqS:GfromQ}. From \Cref{eqS:parity} we can see that essentially the reflection $\alpha\mapsto-\alpha^*$ undoes the transposition $D(\alpha)^{\mathsf T}=D(-\alpha^*)$, so that the criterion measures a parity on $\mathcal T_B(\varrho)$. With the identity pairing $f(\alpha)=\alpha$, the same steps without transposition give $\mathcal L_s^f(\varrho)=\tr(\varrho\,\Pi_{c_\theta})\leq1$, which never detects entanglement. Finally, \cref{eqS:parity} gives the normalization parameter a simple meaning: $s=2\cos^2\theta$ selects the beam-splitter mode whose parity is measured.

\textit{EPR overlap.} For $s=1$ ($\theta=\pi/4$) the parity of $(\hat a-\hat b)/\sqrt2$ is the swap operator of the two modes, i.e., $\Pi_{c_{\pi/4}}=V$ is such that $V\ket{n}\ket{m}=\ket{m}\ket{n}$. Since $\mathcal T_B(V)=\ket{\rm EPR}\bra{\rm EPR}$ with the unnormalized state $\ket{\rm EPR}=\sum_{n\geq0}\ket{nn}$, we obtain
\be\label{eqS:EPR}
\mathcal L_1(\varrho)=\tr[\varrho\,\mathcal T_B(V)]=\bra{\rm EPR}\varrho\ket{\rm EPR}=\sum_{n,m\geq0}\bra{nn}\varrho\ket{mm} .
\ee
Restricting the sum to $n,m<d$ gives $d\,\bra{\psi_+^d}\varrho\ket{\psi_+^d}$, with the maximally entangled state $\ket{\psi_+^d}=d^{-1/2}\sum_{n=0}^{d-1}\ket{nn}$ of the first $d$ Fock states, i.e., the left-hand side of the canonical fidelity witness $d\,\bra{\psi_+^d}\varrho\ket{\psi_+^d}\leq\SN(\varrho)$~\cite{FriisNatPhys19}. The balanced linear criterion is therefore the $d\to\infty$ limit of this fidelity witness, evaluated directly from phase-space data without choosing $d$. \Cref{eqS:EPR} also gives several values at a glance, e.g., $\mathcal L_1(\psi_+^d)=d$ and $\mathcal L_1(\varrho(\bar n)\otimes\varrho(\bar n))=\sum_n p_n^2=1/(2\bar n+1)$ for two thermal states with populations $p_n=\bar n^n/(\bar n+1)^{n+1}$.

\subsection*{C. Relation between the linear and the nonlinear criterion}

For $s=1$ and any relabelling $f$ with $|\det J_f|=1$, the nonlinear criterion is never weaker than the linear one on the same slice,
\be\label{eqS:NLvsL}
\mathcal N^f(\varrho)\geq\mathcal L_1^f(\varrho) ,
\ee
with equality if and only if $X_\varrho^f(\alpha)\geq0$ and $\av{\mathcal Q_A(\alpha)}_\varrho=\av{\mathcal Q_B(f(\alpha))}_\varrho$ for all $\alpha$. Indeed, with the purities $P_j=\tr(\varrho_j^2)$ of the reduced states,
\be
\mathcal L_1^f(\varrho)=\int_{\mathbb C}X_\varrho^f(\alpha)\,\de^2\alpha+\int_{\mathbb C}\av{\mathcal Q_A(\alpha)}_\varrho\av{\mathcal Q_B(f(\alpha))}_\varrho\,\de^2\alpha
\leq\int_{\mathbb C}\bigl|X_\varrho^f(\alpha)\bigr|\,\de^2\alpha+\sqrt{P_AP_B}
\leq\mathcal N^f(\varrho) ,
\ee
where we used the Cauchy--Schwarz inequality first for the integral and then in the form $\sqrt{P_AP_B}+\sqrt{(1-P_A)(1-P_B)}\leq1$. The two criteria thus coincide when the cross-covariance is non-negative and the reduced states have the same characteristic function along the slice, as for the symmetric Gaussian states of Sec.~III, whereas the nonlinear criterion is strictly larger when the reduced characteristic functions differ along the slice, as for the state with asymmetric noise of Sec.~IV. For asymmetric states, however, the linear criterion with an optimized $s\neq1$ can exceed $\mathcal N^f$ (Sec.~III\,D), so that the two criteria are complementary.

\section*{III. Gaussian states}

\subsection*{A. Exact values for two-mode Gaussian states}

For a zero-mean two-mode Gaussian state in the standard form~\eqref{eqS:standardform}, \cref{eqS:chiGauss} gives
\be\label{eqS:chiStd}
\chi_\varrho(\alpha,\beta)=\exp\Bigl\{-\tfrac12\bigl[a|\alpha|^2+b|\beta|^2+2c_1\,\mathrm{Im}\,\alpha\,\mathrm{Im}\,\beta+2c_2\,\mathrm{Re}\,\alpha\,\mathrm{Re}\,\beta\bigr]\Bigr\} .
\ee
Inserting it in $\mathcal L_s(\varrho)=\frac1\pi\int\chi_\varrho(\sqrt s\,\alpha,\sqrt{2-s}\,\alpha^*)\,\de^2\alpha$ (Sec.~II\,B) leaves a Gaussian integral, with the result
{\be\label{eqS:LsGauss}
\mathcal L_s(\varrho)=\frac{2}{\sqrt{(m_s-2c_1\sigma_s)(m_s+2c_2\sigma_s)}} ,
\qquad
m_s=as+b(2-s) ,\quad \sigma_s=\sqrt{s(2-s)} .
\ee}
For the nonlinear criterion with the EPR pairing, the local terms $\av{\mathcal Q_A(\alpha)}_\varrho=e^{-a|\alpha|^2/2}/\sqrt\pi$ and $\av{\mathcal Q_B(-\alpha^*)}_\varrho=e^{-b|\alpha|^2/2}/\sqrt\pi$, together with $c_1>0>c_2$, give $X_\varrho(\alpha)\geq0$, and with $\tr(\varrho_A^2)=1/a$ and $\tr(\varrho_B^2)=1/b$ one finds
{\be\label{eqS:NGauss}
\mathcal N(\varrho)=\mathcal L_1(\varrho)+1-\frac{2}{a+b}-\sqrt{\Bigl(1-\frac1a\Bigr)\Bigl(1-\frac1b\Bigr)} .
\ee}\vspace{0pt}

\textit{Symmetric states.} For $a=b$, \cref{eqS:NGauss} gives $\mathcal N=\mathcal L_1$, in agreement with Sec.~II\,C, and \cref{eqS:LsGauss,eqS:numinus} give
\be\label{eqS:symmetric}
\mathcal L_1(\varrho)=\mathcal N(\varrho)=\frac1\delta ,
\qquad
\delta=\sqrt{(a-c_1)(a+c_2)}=\tilde\nu_- ,
\ee
so that $\SN(\varrho)\geq\lceil1/\delta\rceil$. By \cref{eqS:negativity}, both criteria reach the negativity bound $\|\mathcal T_A(\varrho)\|$ for all entangled symmetric two-mode Gaussian states, and $s=1$ is optimal. For the two-mode squeezed thermal (TMST) state of main-text \cref{eq:TMSTSymDef}, $a=(2\bar n+1)\cosh2\xi$ and $c_1=-c_2=(2\bar n+1)\sinh2\xi$, so that $\delta=(2\bar n+1)e^{-2\xi}$ and the certified Schmidt number is $\lceil e^{2\xi}/(2\bar n+1)\rceil$, as stated in the main text. 
For the two-mode squeezed vacuum $\ket{\psi_\xi}$ ($\bar n=0$), the criteria certify $\SN\geq\lceil e^{2\xi}\rceil$, a finite value although $\SN(\psi_\xi) \rightarrow \infty$. This value is the negativity bound $\|\mathcal T_A(\psi_\xi)\|=e^{2\xi}$, which no criterion bounded by the negativity can exceed, while 
the truncated fidelity witness, whose target coincides with the truncated state, uses the knowledge of its Schmidt spectrum.
This advantage of the fidelity witness is lost already for a slight deviation from purity (main-text \cref{fig:ComparisonFidelityTMSTSym}).

By Sec.~I\,C, the same values are obtained for any two-mode Gaussian state that is brought to the standard form by local symplectic transformations $S_A$ and $S_B$. 

\subsection*{B. Comparison with the entanglement of formation}

For symmetric two-mode Gaussian states, the entanglement of formation (EoF) is known exactly~\cite{GiedkeetAlPRL2003},
\be\label{eqS:EoF}
E_F(\varrho)=h(\delta)=c_+(\delta)\log_2c_+(\delta)-c_-(\delta)\log_2c_-(\delta) ,
\qquad
c_\pm(\delta)=\frac{(1\pm\delta)^2}{4\delta} ,
\ee
with $\delta$ as in \cref{eqS:symmetric}. A state with $\SN(\varrho)\leq r$ is a mixture of pure states with Schmidt rank at most $r$, whose entanglement entropy is at most $\log_2r$. Hence $E_F(\varrho)\leq\log_2\SN(\varrho)$, i.e., the EoF gives the bound $\SN(\varrho)\geq\lceil2^{E_F(\varrho)}\rceil$. Our criteria are always at least as strong.

\begin{proposition}\label{prop:EoF}
For every symmetric two-mode Gaussian entangled state, i.e., $0<\delta<1$, one has $2^{h(\delta)}<1/\delta$, and therefore $\lceil2^{E_F(\varrho)}\rceil\leq\lceil1/\delta\rceil$.
\end{proposition}
\textit{Proof.} The function $u(\delta)=\log_2(1/\delta)-h(\delta)$ vanishes for $\delta\to1$. Using $\de c_\pm/\de\delta=-(1-\delta^2)/(4\delta^2)$, its derivative is
\be
u'(\delta)=-\frac{2\delta-(1-\delta^2)\ln\frac{1+\delta}{1-\delta}}{2\delta^2\ln2}<0 ,
\ee
since $(1-\delta^2)\operatorname{artanh}\delta<\delta$ for $0<\delta<1$, as follows from the power series of $\operatorname{artanh}\delta$. Hence $u(\delta)>0$ for $0<\delta<1$. \qed

The certified integers differ, for instance, for $\delta=0.4$, where our criteria certify $\SN\geq3$ and the EoF only $\SN\geq2$, and for $\delta=0.2$, with $\SN\geq5$ versus $\SN\geq4$. The gap grows with the entanglement: for $\delta\to0$ one finds $2^{h(\delta)}\simeq e/(4\delta)$, so that the bound $1/\delta$ from our criteria exceeds the EoF bound by a factor approaching $4/e\simeq1.47$.

\subsection*{C. Comparison with truncated fidelity witnesses and quadrature-based witnesses}

Main-text \cref{fig:ComparisonFidelityTMSTSym} compares our criteria with truncated fidelity witnesses for the TMST state. \Cref{fig:SMGaussComparisons} shows two further comparisons. In the left panel we consider the two-mode squeezed vacuum mixed with thermal noise,
\be\label{eqS:TMSVmix}
\varrho(p)=p\ket{\psi_\xi}\bra{\psi_\xi}+(1-p)\,\varrho(\bar n)\otimes\varrho(\bar n) ,
\ee
with the thermal state $\varrho(\bar n)$ of main-text \cref{eq:TMSTSymDef}. By linearity and Sec.~III\,A, $\mathcal L_1(\varrho(p))=p\,e^{2\xi}+(1-p)/(2\bar n+1)$, and by Sec.~II\,C the nonlinear criterion gives the same value, since the two reduced states coincide and $X_\varrho(\alpha)\geq0$. We compare this value with the fidelity witness for the two-mode squeezed vacuum truncated to the first $d$ Fock states, main-text \cref{eq:TMSpuretruncated}, evaluated on the state truncated to the same subspace and renormalized. In the right panel we compare the bound $\lceil e^{2\xi}/(2\bar n+1)\rceil$ for the TMST state with the optimized quadrature-based Schmidt-number witnesses of Ref.~\cite{ShahandehOperational2013}. These use quadratic operators $L=\frac12[\omega_1(\hat x_A^2+\hat p_A^2)+\omega_2(\hat x_B^2+\hat p_B^2)+\omega_c(\hat x_A\hat x_B\pm\hat p_A\hat p_B)]$, whose expectation values follow from the CM: a value $\av{L}_\varrho<g_r(L)=\inf_{\SN(\sigma)\leq r}\tr(\sigma L)$, where the threshold is computed numerically, certifies $\SN(\varrho)>r$.

\begin{figure}[h]
\centering
\includegraphics[width=0.48\linewidth]{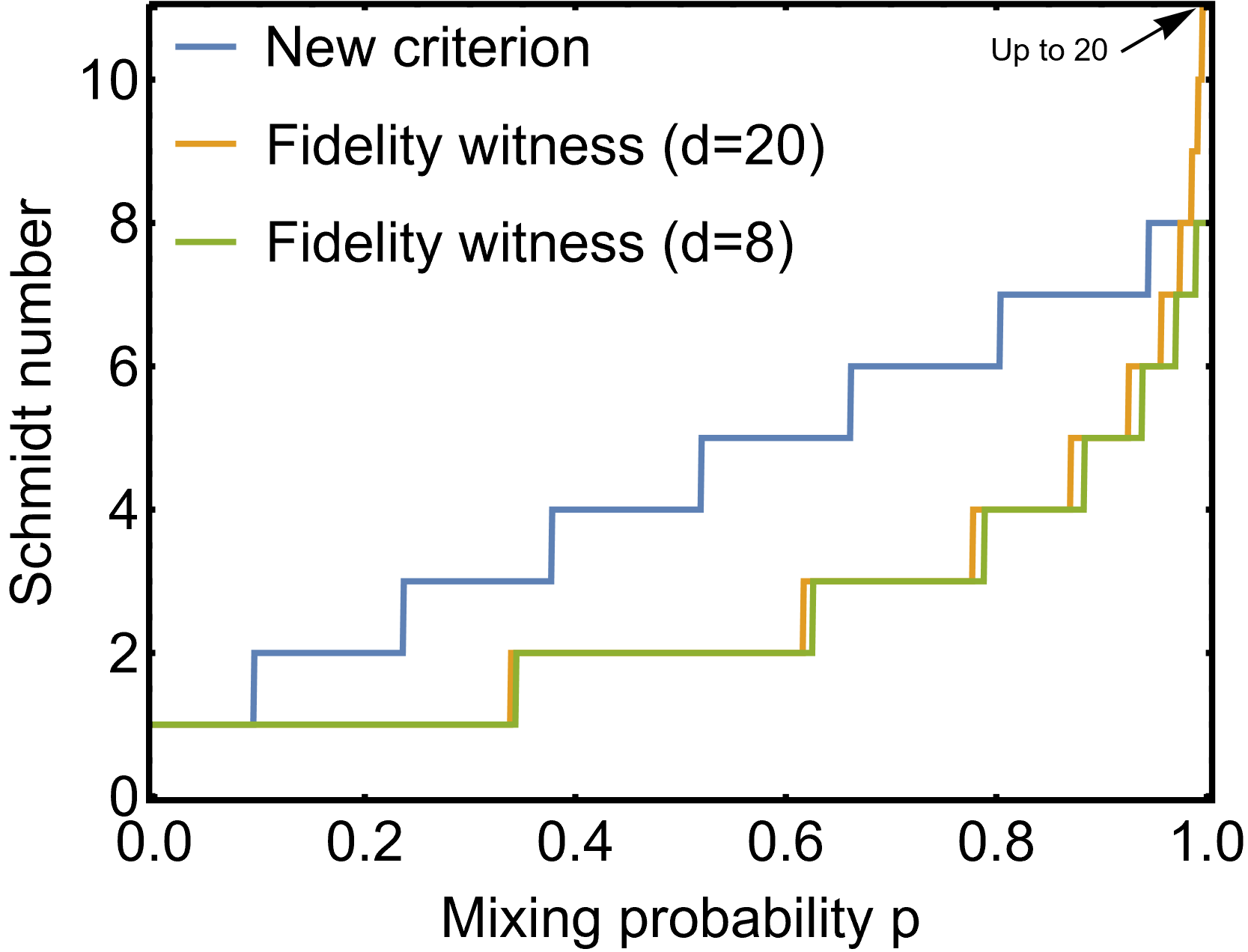}\hfill
\includegraphics[width=0.48\linewidth]{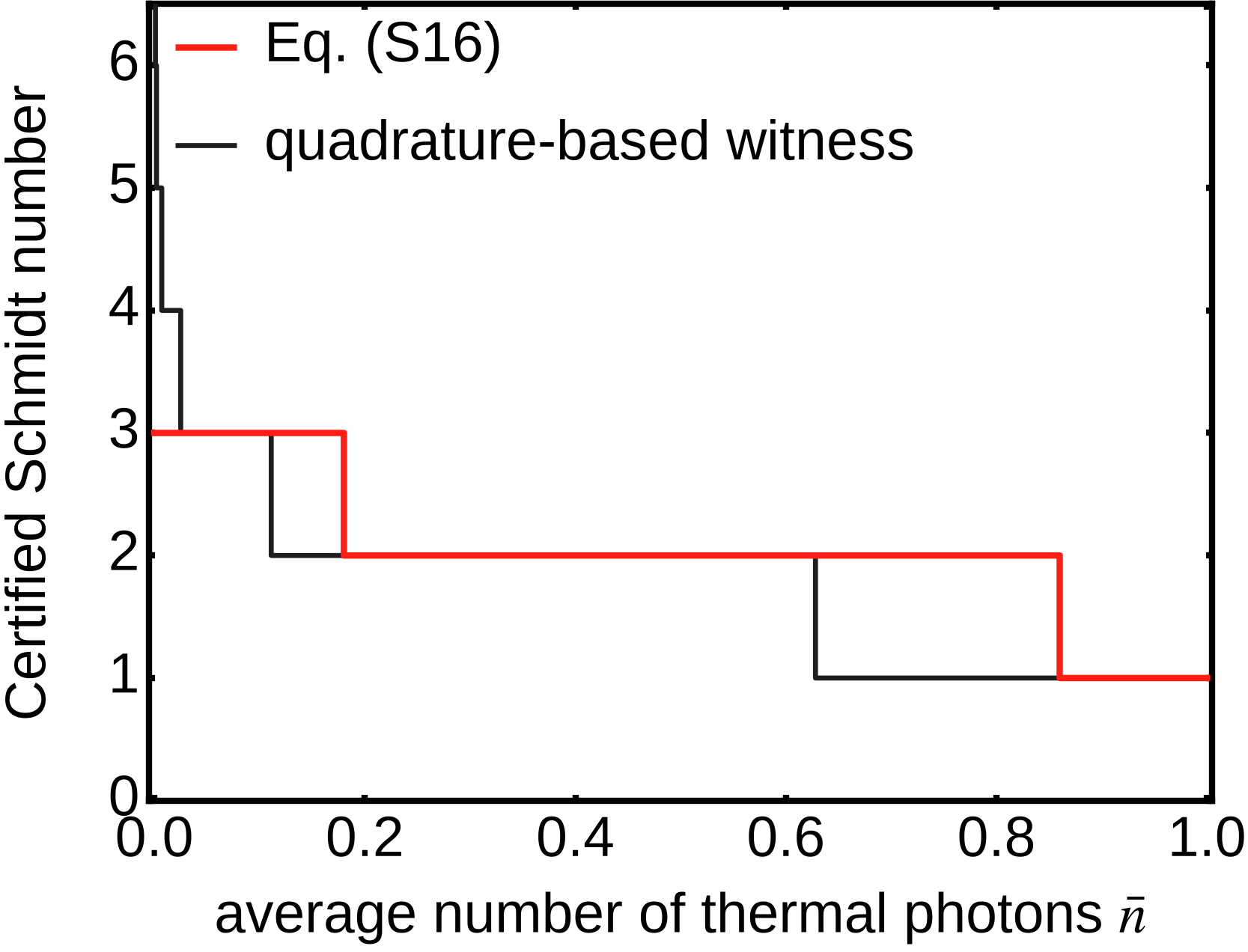}
\caption{Further comparisons for two-mode squeezed states. Left: certified Schmidt number for the state $\varrho(p)$ of \cref{eqS:TMSVmix} with $\xi=1$ and $\bar n=1$, from our criteria and from the fidelity witness for the two-mode squeezed vacuum truncated to $d=20$ Fock states. The fidelity witness certifies a Schmidt number of $20$ only when the state is very pure, whereas for most values of the mixing probability $p$ our criteria give a stronger certification. Right: certified Schmidt number for the TMST state $\varrho_{\rm TMST}(\xi,\bar n)$ of main-text \cref{eq:TMSTSymDef} at $\xi=0.5$, as a function of $\bar n$. The red curve is the bound $\lceil e^{2\xi}/(2\bar n+1)\rceil$ from our criteria, and the black curve is obtained from the optimized quadrature-based witness of Ref.~\cite{ShahandehOperational2013}.}
\label{fig:SMGaussComparisons}
\end{figure}

\subsection*{D. Asymmetric states and the unbalanced normalization}

For asymmetric states, $a\neq b$, the freedom in $s$ becomes important. Let us consider $c_1=-c_2=c>0$, which includes all states obtained from a two-mode squeezed vacuum by phase-insensitive local channels, such as losses, thermal noise or amplification. \Cref{eqS:LsGauss} becomes $\mathcal L_s(\varrho)=2/(m_s-2c\sigma_s)$, which equals $2/(a+b-2c)$ for $s=1$. Writing $m_s-2c\sigma_s=a+b+(a-b)(s-1)-2c\sqrt{1-(s-1)^2}$ and minimizing over $s$ gives
\be\label{eqS:sopt}
\max_{0<s<2}\mathcal L_s(\varrho)=\frac{2}{a+b-\sqrt{(a-b)^2+4c^2}}=\frac1{\tilde\nu_-} ,
\qquad\text{attained at}\quad
s^\ast=1+\frac{b-a}{\sqrt{(a-b)^2+4c^2}} .
\ee
By \cref{eqS:negativity}, the optimal $s$ saturates the first inequality in \cref{eqS:parity}: the linear criterion reaches the negativity bound $\|\mathcal T_A(\varrho)\|$, whereas the balanced value is strictly smaller whenever $a\neq b$. More generally, optimizing $s$ together with linear symplectic relabellings $f_A$ and $f_B$ of the two modes, i.e., using $\mathcal Q_A(\sqrt s\,f_A(\alpha))$ and $\mathcal Q_B(-\sqrt{2-s}\,f_B(\alpha)^*)$, gives $1/\tilde\nu_-$ for all entangled two-mode Gaussian states~\cite{liu2024phasespace}. On Gaussian states, the optimized linear criterion is thus equivalent to the PPT criterion.

\textit{Relative rescaling.} The change of variables $\beta=\sqrt s\,\alpha$ gives $\mathcal L_s(\varrho)=\frac1s\int\av{\mathcal Q_A(\beta)\otimes\mathcal Q_B(-\lambda\beta^*)}_\varrho\,\de^2\beta$ with $\lambda=\sqrt{(2-s)/s}=\tan\theta$. Hence, $s\neq1$ rescales the phase space of $B$ relative to that of $A$, and at the optimum the noisier mode is probed with the smaller displacements, where its characteristic function is still appreciable. A relative rescaling is also possible at $s=1$, with the unit-normalized family $\lambda\,\mathcal Q_B(-\lambda\alpha^*)$ of Sec.~I\,B, as in the witnesses of Refs.~\cite{ZhangDetectingPRL2013,qi2016nonlinear}. Its value, $\lambda\int\av{\mathcal Q_A(\beta)\otimes\mathcal Q_B(-\lambda\beta^*)}_\varrho\,\de^2\beta=\sin(2\theta)\,\mathcal L_s(\varrho)$, is smaller by the factor $\sin2\theta=\sqrt{s(2-s)}\leq1$, and its maximum over $\lambda$, $1/(\sqrt{ab}-c)$~\cite{ZhangDetectingPRL2013}, is smaller than $1/\tilde\nu_-$ whenever $a\neq b$. The unbalanced normalization thus implements the relative rescaling optimally.

\textit{Example.} Consider the TMST state with different thermal occupations of the two modes,
\be\label{eqS:TMSTasym}
\varrho_{\rm TMST}(\xi,\bar n_A,\bar n_B)=U_\xi\bigl[\varrho(\bar n_A)\otimes\varrho(\bar n_B)\bigr]U_\xi^\dagger ,
\ee
with $U_\xi$ as in main-text \cref{eq:TMSTSymDef}. Its CM is in standard form with $a=\Sigma\cosh2\xi+\bar n_A-\bar n_B$, $b=\Sigma\cosh2\xi-\bar n_A+\bar n_B$ and $c_1=-c_2=\Sigma\sinh2\xi$, where $\Sigma=\bar n_A+\bar n_B+1$. For $\bar n_B\to\infty$, the balanced value with the EPR pairing, $\mathcal L_1=e^{2\xi}/\Sigma$, tends to zero, the rescaled balanced value tends to $\sinh2\xi/(2\bar n_A+1)$, and the optimized value
\be
\max_{0<s<2}\mathcal L_s=\frac{1}{\Sigma\cosh2\xi-\sqrt{\Sigma^2\sinh^22\xi+(\bar n_A-\bar n_B)^2}}
\ee
tends to $\cosh2\xi/(2\bar n_A+1)$. For the example of the main text, $\xi=1$, $\bar n_A=0$ and $\bar n_B=3$, one has $a\simeq12.05$, $b\simeq18.05$ and $c\simeq14.51$, and
\be\label{eqS:asymexample}
s=1:\ \mathcal L_1=e^2/4\simeq1.85\ \Rightarrow\ \SN\geq2 ,
\qquad
s^\ast\simeq1.20:\ \mathcal L_{s^\ast}\simeq4.27\ \Rightarrow\ \SN\geq5 .
\ee
Without rescaling, the nonlinear criterion with the EPR pairing gives $\mathcal N\simeq1.85$ from \cref{eqS:NGauss}, and a numerical optimization of the balanced linear criterion over all linear relabellings with $|\det J_f|=1$ gives at most $1.86$, so that both certify only $\SN\geq2$. The rescaled balanced family gives $1/(\sqrt{ab}-c)\simeq4.18$, which here also certifies $\SN\geq5$, and the nonlinear criterion with the same family gives a similar value. For $\bar n_B=5$, instead, the rescaled balanced family certifies only $\SN\geq4$ ($\simeq3.98$ for both criteria), whereas the optimal $s$ certifies $\SN\geq5$ ($\mathcal L_{s^\ast}\simeq4.08$).

\section*{IV. A non-Gaussian example}

To show that the advantage over truncation is not specific to Gaussian states, we consider the maximally entangled state $\ket{\psi_+^d}$ of the first $d$ Fock states mixed with thermal noise. Here the natural benchmark is the fidelity with $\ket{\psi_+^d}$ itself, which detects the pure state optimally. By \cref{eqS:EPR}, the balanced linear criterion is the EPR overlap, which makes the comparison transparent: $\mathcal L_1(\psi_+^d)=d$, and the fidelity witness $d\,\bra{\psi_+^d}\varrho\ket{\psi_+^d}$ is the truncation of $\bra{\rm EPR}\varrho\ket{\rm EPR}$ to the first $d$ Fock states.

\textit{Symmetric noise.} For $\varrho_{\rm sym}(p,\bar n)=p\ket{\psi_+^d}\bra{\psi_+^d}+(1-p)\,\varrho(\bar n)\otimes\varrho(\bar n)$, \cref{eqS:EPR} gives
\be
\mathcal L_1(\varrho_{\rm sym}(p,\bar n))=dp+(1-p)\sum_{n\geq0}p_n^2=dp+\frac{1-p}{2\bar n+1} ,
\ee
with the thermal populations $p_n$ of Sec.~II\,B, whereas the fidelity witness gives $dp+(1-p)\sum_{n<d}p_n^2=dp+(1-p)\bigl[1-\bigl(\tfrac{\bar n}{\bar n+1}\bigr)^{2d}\bigr]/(2\bar n+1)$. Our bound is larger by the contribution of the Fock states beyond the truncation, which is small in this case.

\textit{Asymmetric noise.} For the state of main-text \cref{eq:MESAsymmetricNoise}, $\varrho(p,\bar n)=p\ket{\psi_+^d}\bra{\psi_+^d}+(1-p)\,\varrho(\bar n)\otimes\ket{0}\bra{0}$, only the vacuum component of the noise contributes to the EPR overlap, and $\mathcal L_1$ coincides with the fidelity witness evaluated without renormalization, $dp+(1-p)/(\bar n+1)$. The two reduced states, however, now have different purities, and by Sec.~II\,C the nonlinear criterion is strictly larger than $\mathcal L_1$ for $p<1$. \Cref{fig:MESwithasymmetricnoise} shows the resulting certified Schmidt numbers for $d=5$ and $\bar n=0.5$. As stated in the main text, the fidelity witness is evaluated after truncating the noise to the first $d$ Fock states and renormalizing it, which slightly increases its value. The nonlinear criterion certifies a Schmidt number at least as large as the fidelity witness for all $p$, and a larger one in several ranges of $p$.

\begin{figure}[h]
\centering
\includegraphics[width=0.55\linewidth]{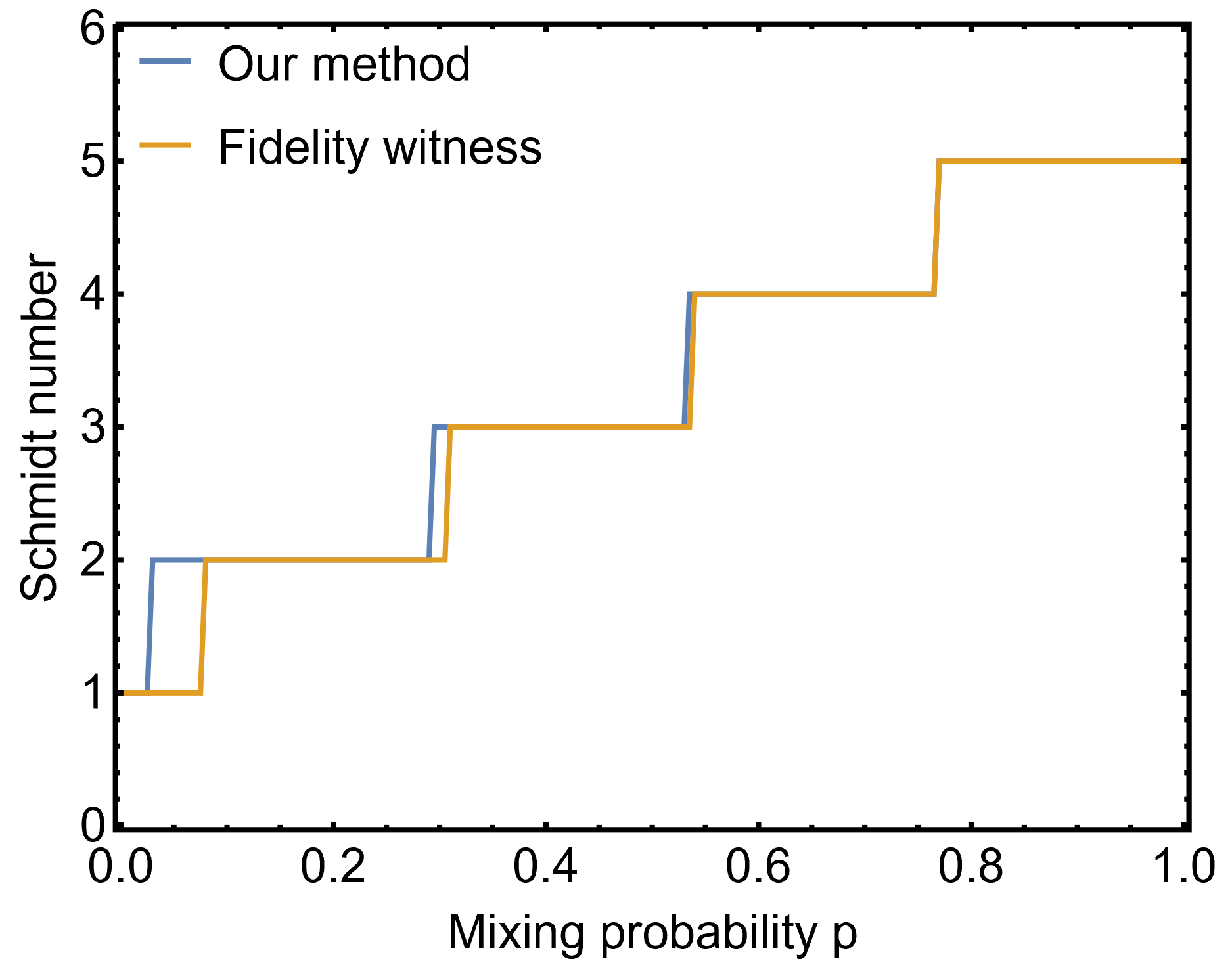}
\caption{Certified Schmidt number $\SN_{\rm cert}$ as a function of the mixing probability $p$ for the state of main-text \cref{eq:MESAsymmetricNoise} with $d=5$ and $\bar n=0.5$, from the nonlinear criterion, main-text \cref{eq:CVSNCriterion}, and from the fidelity witness with respect to the pure $d$-dimensional maximally entangled state.}
\label{fig:MESwithasymmetricnoise}
\end{figure}

\section*{V. Multiple copies of a Gaussian bound-entangled state}

Here we show that the multimode nonlinear criterion~\eqref{eqS:nonlinearM} certifies Schmidt numbers that grow exponentially with the number of copies of a Gaussian state that remains PPT, in analogy with the finite-dimensional constructions of Refs.~\cite{HuberLamiLancienMuellerHermes2018,KrebsMariPRL2024,KrebsGachechiladzePRA2026}.

\textit{The state.} We consider the zero-mean $2\times2$-mode Gaussian states of Refs.~\cite{WernerWolfPRL2001,ZhangDetectingPRL2013}, in which party $A$ holds the modes $A_1,A_2$ and party $B$ the modes $B_1,B_2$. In the quadrature ordering $(\hat x_{A_1},\hat p_{A_1},\hat x_{A_2},\hat p_{A_2},\hat x_{B_1},\hat p_{B_1},\hat x_{B_2},\hat p_{B_2})$, the CM of the state $\varrho(a,c)$ is
\be\label{eqS:WWcm}
\Gamma(a,c)=\begin{pmatrix}aI_4&cR\\cR&aI_4\end{pmatrix},
\qquad
R=\begin{pmatrix}1&0&0&0\\0&0&0&-1\\0&0&-1&0\\0&-1&0&0\end{pmatrix},
\ee
with $a\geq1$ and $c>0$. The local blocks are already in the standard form~\eqref{eqS:Williamson}: the reduced states are products of thermal states with $\bar n=(a-1)/2$ photons per mode. $\Gamma(a,c)$ is a valid CM, $\Gamma(a,c)+i\Omega_4\geq0$, if and only if
\be\label{eqS:cmax}
c\leq c_{\max}(a)=\sqrt{a^2-\sqrt{2a^2-1}}=\sqrt{a^2-\tfrac12}-\tfrac{1}{\sqrt2} .
\ee
The state is PPT for all $a$ and $c$~\cite{WernerWolfPRL2001,ZhangDetectingPRL2013}: partial transposition maps $\Gamma(a,c)$ to the CM obtained from it by a local phase rotation by $\pi$ of the modes $A_2$ and $B_2$, so that $\mathcal T_B(\varrho(a,c))$ is unitarily equivalent to $\varrho(a,c)$ and hence positive. Moreover, $\varrho(a,c)$ is separable for $c\leq a-1$~\cite{WernerWolfPRL2001} and entangled for $c>a-1$, where its realignment norm $(a-c)^{-2}$ exceeds one~\cite{ZhangDetectingPRL2013}. For $a-1<c\leq c_{\max}(a)$, a nonempty interval for $a>1$, $\varrho(a,c)$ is therefore bound entangled. Gaussian bound-entangled states of light have been prepared and detected experimentally~\cite{DiGuglielmoExperimental2011}. The $n$-copy state $\varrho(a,c)^{\otimes n}$, with $2n$ modes in each party, remains PPT, since $\mathcal T_{B}(\varrho^{\otimes n})=[\mathcal T_B(\varrho)]^{\otimes n}\geq0$, and no entanglement can be distilled from it~\cite{HorodeckiLewensteinBoundEntanglement}.

\textit{The slice.} We evaluate \cref{eqS:nonlinearM} with $M=2n$ on the linear slice $f(\vec y)=F^{\oplus n}\vec y$, where $\vec y=(\vec y_1,\dots,\vec y_n)\in\mathbb R^{4n}$ and $\vec y_k\in\mathbb R^4$ collects the labels of the modes $A_1,A_2$ of the $k$-th copy, with
\be\label{eqS:F}
F=\begin{pmatrix}0&0&-1&0\\0&1&0&0\\-1&0&0&0\\0&0&0&-1\end{pmatrix}.
\ee
$F$ is symmetric and orthogonal, with $F^2=I_4$ and $\det F=1$, so that $|\det J_f|=1$. It is worth noting that this transformation is neither symplectic nor anti-symplectic, $F^{\mathsf T}\Omega_2F\neq\pm\Omega_2$, so that this slice is not equivalent to the EPR pairing up to local Gaussian unitaries.

\begin{proposition}\label{prop:multicopy}
Let $a\geq1$, $0<c\leq c_{\max}(a)$ and $n\geq1$. On the slice $f(\vec y)=F^{\oplus n}\vec y$, the nonlinear criterion~\eqref{eqS:nonlinearM} for $\varrho(a,c)^{\otimes n}$ takes the value
\be\label{eqS:multicopy}
\mathcal N^f\bigl(\varrho(a,c)^{\otimes n}\bigr)=\frac{1}{(a-c)^{2n}} ,
\qquad\text{so that}\qquad
\SN\bigl(\varrho(a,c)^{\otimes n}\bigr)\geq\left\lceil\frac{1}{(a-c)^{2n}}\right\rceil .
\ee
Moreover, $(a-c)^{-2}<2$ for all such $a$ and $c$, while $(a-c_{\max}(a))^{-2}\to2$ for $a\to\infty$.
\end{proposition}
\textit{Proof.} The characteristic function of $\varrho(a,c)^{\otimes n}$ is the product of those of the copies. From \cref{eqS:chiGauss,eqS:WWcm,eqS:F} one finds, on the slice and on the local subspaces,
\be\label{eqS:chislice}
\chi_{\varrho(a,c)^{\otimes n}}\bigl(\vec y,-F^{\oplus n}\vec y\bigr)=e^{-(a-c)\|\vec y\|^2} ,
\qquad
\chi_{\varrho_A(a,c)^{\otimes n}}(\vec y)=\chi_{\varrho_B(a,c)^{\otimes n}}\bigl(F^{\oplus n}\vec y\bigr)=e^{-a\|\vec y\|^2/2} .
\ee
These functions are real and even, so that, by the multimode form of \cref{eqS:Qmean,eqS:Qcorr}, $X^f(\vec y)=\pi^{-2n}\bigl[e^{-(a-c)\|\vec y\|^2}-e^{-a\|\vec y\|^2}\bigr]\geq0$ and $\tr[(\varrho_A^{\otimes n})^2]=\tr[(\varrho_B^{\otimes n})^2]=a^{-2n}$. With $\int_{\mathbb R^{4n}}e^{-\kappa\|\vec y\|^2}\de^{4n}\vec y=(\pi/\kappa)^{2n}$, the left-hand side of \cref{eqS:nonlinearM} becomes $(a-c)^{-2n}-a^{-2n}-(1-a^{-2n})+1=(a-c)^{-2n}$. Finally, \cref{eqS:cmax} gives $a-c_{\max}(a)=a-\sqrt{a^2-\frac12}+\frac{1}{\sqrt2}$, which is larger than $1/\sqrt2$ and tends to $1/\sqrt2$ for $a\to\infty$. \qed

For $n=1$, the value $(a-c)^{-2}$ is the realignment norm $\|\mathcal R(\varrho(a,c))\|$~\cite{ZhangDetectingPRL2013}. Since the realignment norm is multiplicative under tensor products, on this slice the criterion reaches the realignment bound $\|\mathcal R(\varrho^{\otimes n})\|\leq\SN(\varrho^{\otimes n})$ for all $n$, while requiring only correlations on a $4n$-dimensional slice of the $8n$-dimensional phase space. The multimode form of the balanced linear criterion~\eqref{eqS:linear} gives the same value on this slice.

\textit{Exponential growth.} For $a-1<c\leq c_{\max}(a)$, the base $(a-c)^{-2}$ is larger than one and, by Proposition~\ref{prop:multicopy}, smaller than two: a single copy certifies at most $\SN\geq2$, but for any $v<2$ there are parameters $(a,c)$ such that $\SN(\varrho(a,c)^{\otimes n})\geq\lceil v^n\rceil$ for all $n$. \Cref{tab:multicopy} lists the certified Schmidt numbers at $c=c_{\max}(a)$. For instance, two copies certify $\SN\geq3$ already for $a=2$, i.e., for half a thermal photon per mode, and for $a=40$ three copies certify $\SN\geq8$. If the copies are prepared independently, the characteristic functions entering \cref{eqS:multicopy} are products of single-copy functions on the four-dimensional slice $\{(\vec y_k,-F\vec y_k)\}$ and on the local subspaces, so that the measurement cost is that of a single copy.

\begin{table}[h]

\caption{Certified Schmidt number $\lceil(a-c)^{-2n}\rceil$ of $n$ copies of the state $\varrho(a,c)$ at the boundary $c=c_{\max}(a)$, for different values of the local parameter $a$, with $\bar n=(a-1)/2$ thermal photons per mode.}
\label{tab:multicopy}
\begin{ruledtabular}
\begin{tabular}{cccccccc}
$a$ & $\bar n$ & $c_{\max}(a)$ & $(a-c)^{-2}$ & $n=1$ & $n=2$ & $n=3$ & $n=4$\\
\hline
$2$ & $0.5$ & $1.1637$ & $1.4299$ & $2$ & $3$ & $3$ & $5$\\
$5$ & $2$ & $4.2426$ & $1.7434$ & $2$ & $4$ & $6$ & $10$\\
$10$ & $4.5$ & $9.2679$ & $1.8656$ & $2$ & $4$ & $7$ & $13$\\
$40$ & $19.5$ & $39.2866$ & $1.9651$ & $2$ & $4$ & $8$ & $15$\\
\end{tabular}
\end{ruledtabular}
\end{table}

\textit{No fidelity witness detects these states.} For a PPT state $\varrho$ and a pure state $\ket\psi$ with largest squared Schmidt coefficient $\lambda_1$, one has $\bra\psi\varrho\ket\psi=\tr[\mathcal T_B(\varrho)\,\mathcal T_B(\ket\psi\bra\psi)]\leq\lambda_1$, because $\mathcal T_B(\varrho)$ is a density operator and the largest eigenvalue of $\mathcal T_B(\ket\psi\bra\psi)$ is $\lambda_1$. Certifying $\SN\geq2$ with a fidelity witness, however, requires $\bra\psi\varrho\ket\psi>\lambda_1$. Truncating the local Hilbert spaces does not help, since local projections onto subspaces spanned by Fock states are invariant under transposition in the Fock basis and therefore preserve the PPT property.

\section*{VI. Experimental implementation}

The criteria require the local expectation values $\av{\mathcal Q(\alpha)}$ and the correlations $\av{\mathcal Q_A(\alpha)\otimes\mathcal Q_B(f(\alpha))}$ on a two-dimensional slice, i.e., the two-mode characteristic function along a plane rather than in the full four-dimensional phase space\footnote{Since the characteristic function satisfies $\chi(-\alpha)=\chi(\alpha)^*$, half of the plane is sufficient.}. For data that already sample the required points, a new relabelling $f$ amounts to relabelling the data, and otherwise it requires another two-dimensional scan, but never full two-mode tomography. In all examples of the main text and of this SM, it is sufficient to apply $f$ to mode $B$ only.

\subsection*{A. Ancilla-based measurements}

In systems with a Jaynes--Cummings-type coupling of the bosonic mode to a two-level ancilla, the characteristic function can be measured with conditional displacements and a qubit readout~\cite{WilkensPRA91,KimPRA98,asadian_probing_2014,johnson_sensing_2015,campagne-ibarcq_quantum_2020,FluhmannPRL20}, and closely related protocols measure the Wigner function~\cite{LutterbachPRL97}. These techniques are routinely used in trapped ions~\cite{leibfried_experimental_1996,johnson_sensing_2015,FluhmannPRL20}, cavity and circuit quantum electrodynamics~\cite{bertet_direct_2002,vlastakis_deterministically_2013,campagne-ibarcq_quantum_2020,eickbusch_fast_2022,diringer_conditional-not_2024,valadares_-demand_2024}, and circuit quantum acoustodynamics~\cite{satzinger_quantum_2018,von_lupke_parity_2022}, and have been extended to several modes both for the characteristic function~\cite{LinPRL24} and for the Wigner function~\cite{ChenScience2016,gao_entanglement_2019,HeNatCommun2024,JeonPRXQ2025}.

\Cref{fig:Circuit} shows two schemes for our criteria. Both use qubit $\pi/2$ rotations with adjustable phase,
$R_{\pi/2}(\varphi)=\frac{1}{\sqrt2}\bigl(\begin{smallmatrix}1&e^{i\varphi}\\-e^{-i\varphi}&1\end{smallmatrix}\bigr)$,
conditional displacements $U=\ket0\bra0\otimes\id+\ket1\bra1\otimes D(\alpha)$, and a final projective readout of the ancilla~\cite{AsadianHeisenbergWeyl2016,LinPRL24}. Each scheme realizes a two-outcome measurement $\{E_0,E_1\}$ with Kraus operators $M_j$, $E_j=M_j^\dagger M_j$. We assign the value $\boldsymbol s=+1$ to the outcome $j=1$ and $\boldsymbol s=-1$ to $j=0$, so that the mean outcome is $\av{\boldsymbol s}=\av{E_1-E_0}$, which is directly proportional to the CVLOOs:
\begin{itemize}
\item[(a)] \textit{One ancilla per mode.} For each mode, the Kraus operators $M_0=\frac12[\id-e^{i\pi/4}D(\alpha)]$ and $M_1=\frac12[-\id-e^{i\pi/4}D(\alpha)]$ give $E_1-E_0=\frac12[e^{i\pi/4}D(\alpha)+e^{-i\pi/4}D^\dagger(\alpha)]=\sqrt{\pi/2}\,\mathcal Q(\alpha)$. With the outcomes $\boldsymbol s^{(A)},\boldsymbol s^{(B)}\in\{\pm1\}$ of the two ancillas and sample means denoted by a bar, one thus obtains $\av{\mathcal Q(\alpha)}=\sqrt{2/\pi}\;\overline{\boldsymbol s}$ for each mode and
\be
\hat X_\varrho(\alpha)=\frac2\pi\Bigl(\overline{\boldsymbol s^{(A)}\boldsymbol s^{(B)}}-\overline{\boldsymbol s^{(A)}}\;\overline{\boldsymbol s^{(B)}}\Bigr)
\ee
for the cross-covariance, from the same data.
\item[(b)] \textit{One ancilla for both modes.} A single ancilla coupled sequentially to the two modes realizes, with $D_A=D(\alpha)\otimes\id$ and $D_B=\id\otimes D(f(\alpha))$, the Kraus operators
\be
M_0=\tfrac{1}{2\sqrt2}\bigl[\id-ie^{i\pi/4}D_A-e^{-i\pi/4}D_B-iD_AD_B\bigr],
\qquad
M_1=\tfrac{1}{2\sqrt2}\bigl[-\id+ie^{i\pi/4}D_A-e^{-i\pi/4}D_B-iD_AD_B\bigr],
\ee
which give $E_1-E_0=\frac14\bigl[iD_AD_B+D_AD_B^\dagger+D_A^\dagger D_B-iD_A^\dagger D_B^\dagger\bigr]=\frac\pi2\,\mathcal Q_A(\alpha)\otimes\mathcal Q_B(f(\alpha))$, i.e., $\av{\mathcal Q_A(\alpha)\otimes\mathcal Q_B(f(\alpha))}=\frac2\pi\av{\boldsymbol s}$. Replacing the conditional displacement of mode $B$ with the identity gives instead $E_1-E_0=\frac{\sqrt\pi}{2}\mathcal Q_A(\alpha)$, and similarly for mode $A$, which provides the local terms of $X_\varrho(\alpha)$. A related scheme was used in Ref.~\cite{ChenScience2016}.
\end{itemize}

\begin{figure}[h]
\centering
\includegraphics[width=0.6\linewidth]{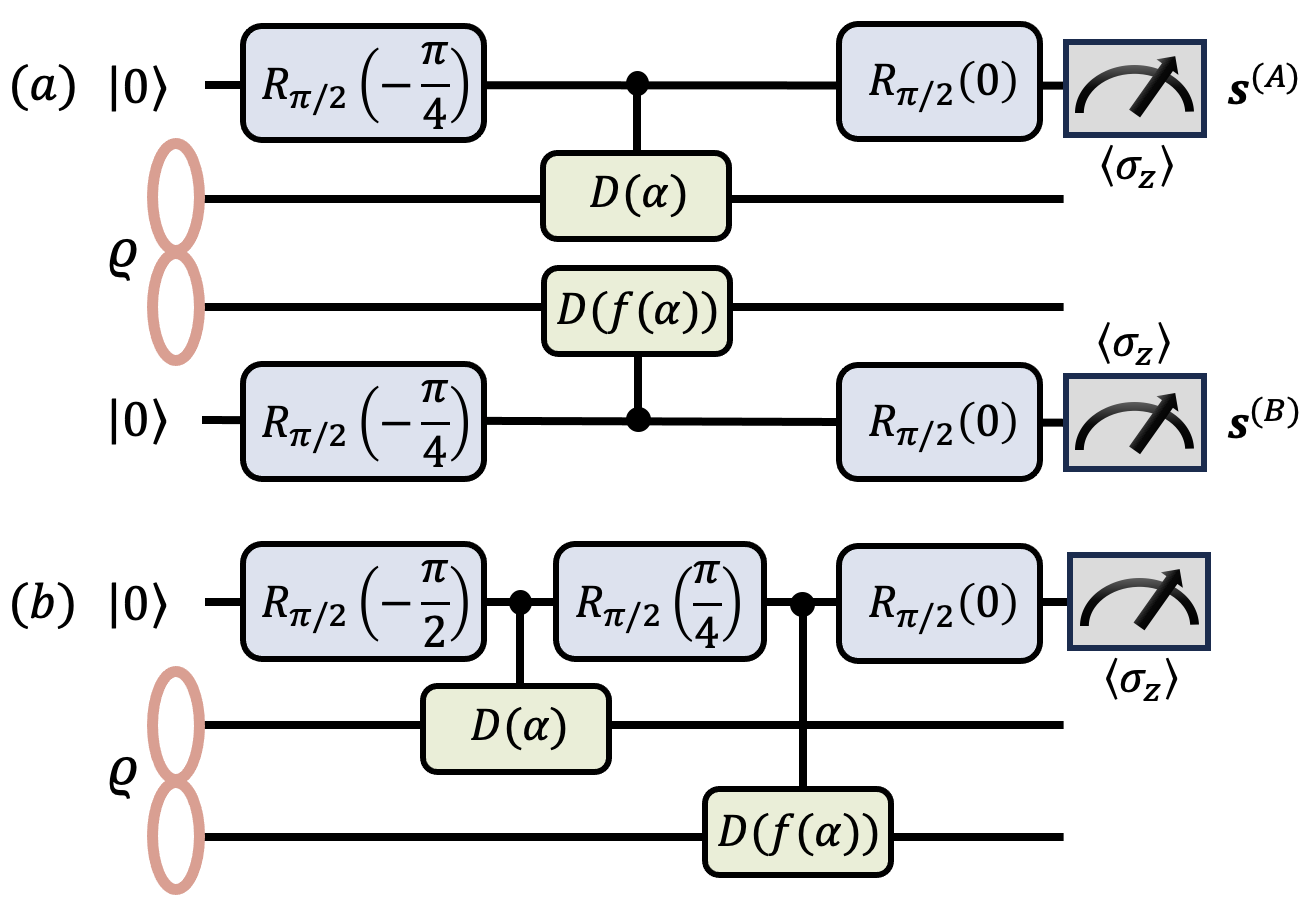}
\caption{Experimental implementations of main-text \cref{eq:CCNR_PPT_CVLOO_r,eq:CVSNCriterion}.
(a) Each CV mode of $\varrho$ is coupled to its own ancillary readout qubit, and the correlations between the CVLOOs $\mathcal{Q}(\alpha)$ are reconstructed from the binary outcomes $s\in\{\pm1\}$.
(b) Both CV modes interact sequentially with a single ancillary qubit, such that $\av{\mathcal{Q}_A(\alpha)\otimes\mathcal{Q}_B(f(\alpha))}$ is obtained from the measured expectation value $\av{\sigma_z}$.}
\label{fig:Circuit}
\end{figure}

\subsection*{B. Homodyne detection}

For optical fields, where displacement and parity measurements are less common, the same slice can be accessed with two balanced homodyne detectors. Writing $\alpha=|\alpha|e^{i\varphi}$, \cref{eqS:displacement} gives $D(\alpha)=\exp(i|\alpha|\hat X_{\varphi-\pi/2})$, where $\hat X_\vartheta=\hat x\cos\vartheta+\hat p\sin\vartheta$ is the quadrature measured at local-oscillator phase $\vartheta$. For a linear relabelling with $f(|\alpha|e^{i\varphi})=|\alpha|\kappa_\varphi e^{i\varphi_B}$, where $\kappa_\varphi$ and $\varphi_B$ depend on $\varphi$, one then obtains
\be
\chi_\varrho\bigl(\alpha,\pm f(\alpha)\bigr)=\Bigl\langle\exp\bigl[i|\alpha|\bigl(\hat X^A_{\varphi-\pi/2}\pm\kappa_\varphi\hat X^B_{\varphi_B-\pi/2}\bigr)\bigr]\Bigr\rangle_\varrho ,
\ee
For fixed $\varphi$, the measured joint distribution $p_\varphi(x_A,x_B)$ of the two homodyne outcomes thus determines the characteristic function on the entire line $\alpha=|\alpha|e^{i\varphi}$, $|\alpha|\in\mathbb R$, through a one-dimensional Fourier transform,
\be\label{eqS:homodyneFT}
\chi_\varrho\bigl(\alpha,\pm f(\alpha)\bigr)=\int_{\mathbb R^2}p_\varphi(x_A,x_B)\,e^{i|\alpha|(x_A\pm\kappa_\varphi x_B)}\,\de x_A\,\de x_B ,
\ee
as in homodyne tomography~\cite{LvovskyContinuous2009}, and its marginals give $\chi_\varrho(\alpha,0)$ and $\chi_\varrho(0,f(\alpha))$ in the same way, so that \cref{eqS:Qmean,eqS:Qcorr} yield $\av{\mathcal Q_A(\alpha)}$, $\av{\mathcal Q_B(f(\alpha))}$, $\av{\mathcal Q_A(\alpha)\otimes\mathcal Q_B(f(\alpha))}$ and the local purities from the same data. Only the angle $\varphi\in[0,\pi)$ has to be scanned, while the integrals over $\alpha$ in \cref{eqS:linear,eqS:nonlinear} are carried out in polar coordinates, $\de^2\alpha=|\alpha|\,\de|\alpha|\,\de\varphi$, on the Fourier-transformed data. A one-parameter scan of correlated local-oscillator phases therefore gives access to all quantities entering the criteria. For the EPR pairing $f(\alpha)=-\alpha^*=|\alpha|e^{i(\pi-\varphi)}$, $\kappa_\varphi=1$ and $\varphi_B=\pi-\varphi$, so that $\vartheta_B=-\vartheta_A$: the two local-oscillator phases are scanned in opposite directions, and for $\vartheta_A=0$ and $\vartheta_A=\pi/2$ the combinations $\hat X^A_{\vartheta_A}\mp\hat X^B_{-\vartheta_A}$ are the EPR variables $\hat x_A\mp\hat x_B$ and $\hat p_A\pm\hat p_B$.

\clearpage
\end{widetext}

\end{document}